\documentclass[12pt]{article}

\usepackage[utf8]{inputenc}
\usepackage{jheppub}
\usepackage{epsfig}
\usepackage{bbm}
\usepackage{amsmath}
\usepackage{amsfonts}
\usepackage{amssymb}
\usepackage{mathtools}
\usepackage{dsfont}
\usepackage{bm}
\usepackage{graphicx}
\usepackage{longtable}
\usepackage{pdflscape}
\usepackage{subcaption}
\usepackage{color}
\usepackage{psfrag}
\usepackage[capitalise]{cleveref}
\usepackage{hyperref}
\usepackage{tikz}
\usetikzlibrary{positioning}
\usepackage{pgfplots}
\usepackage{subfiles}
\usepackage{longtable}
\usepackage{tabularx}
\usepackage{ltablex}
\usepackage{booktabs}
\usepackage[export]{adjustbox}
\usepackage{listings}
\usepackage{multirow} 
\usepackage{cancel,slashed}
\usepackage{bbm}
\usepackage{graphicx}
\usepackage{latexsym}
\usepackage{color}
\usepackage{transparent}
\usepackage{mathtools}
\usepackage{array}
\usepackage{makecell}
\usepackage{graphics,psfrag}
\usepackage{placeins}
\usepackage{nowidow}
\usepackage{listings}
\usepackage[normalem]{ulem}
\usepackage{environ}

\pgfplotsset{compat=1.15}

\makeatletter
\def\@xfootnote[#1]{%
	\protected@xdef\@thefnmark{#1}%
	\@footnotemark\@footnotetext}
\makeatother

\newcommand{\de}{\partial}

\newcommand{\PP}{\mathbb{P}}

\newcommand{\ZZ}{\mathbb{Z}}
\newcommand{\ID}{\mathds{1}}

\newcommand{\pderr}[1]{\frac{\de}{\de #1}}

\newcommand{\coma}{\, , \quad}
\newcommand{\fstop}{\, .}

\newcommand{\AdS}{\text{AdS}}
\newcommand{\dS}{\text{dS}}
\newcommand{\dP}{\text{dP}}

\renewcommand{\epsilon}{\varepsilon}

\makeatletter
\newsavebox{\measure@tikzpicture}
\NewEnviron{scaletikzpicturetowidth}[1]{%
  \def\tikz@width{#1}%
  \def\tikzscale{1}\begin{lrbox}{\measure@tikzpicture}%
  \BODY
  \end{lrbox}%
  \pgfmathparse{#1/\wd\measure@tikzpicture}%
  \edef\tikzscale{\pgfmathresult}%
  \BODY
}
\makeatother

\def\im{{\rm Im \,}}
\def\re{{\rm Re \,}}

\catcode`\@=11   
\newdimen\@rotdimen
\newbox\@rotbox  

\def\@vspec#1{\special{ps:#1}}
\def\@rotstart#1{\@vspec{gsave currentpoint currentpoint translate
		#1 neg exch neg exch translate}}
\def\@rotfinish{\@vspec{currentpoint grestore moveto}}
\def\@rotr#1{\@rotdimen=\ht#1\advance\@rotdimen by\dp#1%
	\hbox to\@rotdimen{\hskip\ht#1\vbox to\wd#1{\@rotstart{90 rotate}%
			\box#1\vss}\hss}\@rotfinish}
\def\@rotl#1{\@rotdimen=\ht#1\advance\@rotdimen by\dp#1%
	\hbox to\@rotdimen{\vbox to\wd#1{\vskip\wd#1\@rotstart{270 rotate}%
			\box#1\vss}\hss}\@rotfinish}%
\def\@rotu#1{\@rotdimen=\ht#1\advance\@rotdimen by\dp#1%
	\hbox to\wd#1{\hskip\wd#1\vbox to\@rotdimen{\vskip\@rotdimen
			\@rotstart{-1 dup scale}\box#1\vss}\hss}\@rotfinish}%
\def\@rotf#1{\hbox to\wd#1{\hskip\wd#1\@rotstart{-1 1 scale}%
		\box#1\hss}\@rotfinish}%
\def\rotate{\@ifnextchar[{\@rotate}{\@rotate[l]}}
\def\@rotate[#1]#2{\setbox\@rotbox=\hbox{#2}\@nameuse{@rot#1}\@rotbox}

\catcode`\@=12

\hypersetup{
	pdftitle={Gopakumar-Vafa Hierarchies in Winding Inflation and Uplifts},    
	pdfauthor={\textcopyright\ Federico Carta, Alessandro Mininno, Nicole Righi, Alexander Westphal},     
	pdfsubject={HEP},   
	pdfcreator={pdfLaTex},   
	pdfproducer={LaTex}, 
	pdfkeywords={},
}

\allowdisplaybreaks[1] 

\begin{document}
	\pagestyle{plain}

	\makeatletter
	\@addtoreset{equation}{section}
	\makeatother
	\renewcommand{\theequation}{\thesection.\arabic{equation}}
	\pagestyle{empty}
\rightline{IFT-UAM/CSIC-21-3}
\rightline{DESY 21-007}
\vspace{3cm}

\begin{center}
	\LARGE{\bf Gopakumar-Vafa Hierarchies in Winding Inflation and Uplifts}\\
	\large{Federico Carta,\textsuperscript{1} Alessandro Mininno,\textsuperscript{2} Nicole Righi,\textsuperscript{3} Alexander Westphal\textsuperscript{3}\\[4mm]}
	\footnotesize{\textsuperscript{1}Department of Mathematical Sciences, Durham University,\\
	Durham, DH$1$ $3$LE, United Kingdom\\
		\textsuperscript{2}Instituto de F\'{\i}sica Te\'orica IFT-UAM/CSIC,\\
		C/ Nicol\'as Cabrera 13-15, 
		Campus de Cantoblanco, 28049 Madrid, Spain\\
	\textsuperscript{3}Deutches Electronen-Synchrotron, DESY,\\ Notkestra\ss e 85, 22607 Hamburg, Germany}\\
	\footnotesize{\href{mailto:federico.carta@durham.ac.uk}{federico.carta@durham.ac.uk}, \href{mailto:alessandro.mininno@uam.es}{alessandro.mininno@uam.es}, \href{mailto:nicole.righi@desy.de}{nicole.righi@desy.de}, \href{mailto:alexander.westphal@desy.de}{alexander.westphal@desy.de}}

\vspace*{4mm}

\small{\bf Abstract} 
\\[4mm]
\end{center}
\begin{center}
\begin{minipage}[h]{\textwidth}
We propose a combined mechanism to realize both winding inflation and de Sitter uplifts. We realize the necessary structure of competing terms in the scalar potential not via tuning the vacuum expectation values of the complex structure moduli, but by a hierarchy of the Gopakumar-Vafa invariants of the underlying Calabi-Yau threefold. To show that Calabi-Yau threefolds with the prescribed hierarchy actually exist, we explicitly create a database of all the genus $0$ Gopakumar-Vafa invariants up to total degree $10$ for all the complete intersection Calabi-Yau's up to Picard number $9$. As a side product, we also identify all the redundancies present in the CICY list, up to Picard number $13$. Both databases can be accessed at this \href{https://www.desy.de/~westphal/GV_CICY_webpage/GVInvariants.html}{link}.
\end{minipage}
\end{center}

	\newpage
	\setcounter{page}{1}
	\pagestyle{plain}
	\renewcommand{\thefootnote}{\arabic{footnote}}
	\setcounter{footnote}{0}
	
	\tableofcontents
	

\section{Introduction}	
\label{sec:Introduction}

An ongoing series of observational cosmological probes, among them cosmic microwave background (CMB) measurements~\cite{Hinshaw:2012aka,Akrami:2018odb,Ade:2018gkx}, type IA supernova data (see e.g.~\cite{Scolnic:2017caz}), large-scale structure (LSS) surveys (see e.g. the very recent results of \cite{Abbott:2017wau}), and baryon acoustic oscillation (BAO) measurements (see e.g.~\cite{deMattia:2020fkb}), has so far provided increasing evidence for the $\Lambda$CDM cosmological standard model. In particular, this includes support for a concurrent late-time accelerating expansion of the universe compatible with a description by de Sitter (dS) space with a very small positive cosmological constant (c.c.) and a very early epoch of extremely rapid exponential expansion called inflation.

This observational background provides the motivation for continued efforts to search for vacuum solutions (``vacua") of string theory as a candidate theory of quantum gravity which can realize both controlled dS vacua and an observationally viable epoch of slow-roll inflation. In many cases, constructing such string vacua involves stabilizing all moduli scalar fields, and stringy p-form axion fields using fluxes and orientifold planes up to typically one or two scalar field directions left massless and ``flat" at leading order. For these remaining few flat directions a scalar potential arises from taking  into account non-perturbative quantum corrections, in particular if these flat directions are axions for which perturbative corrections are absent.

It is in this context where the study of beautiful mathematical objects of Calabi-Yau (CY) quantum geometry such as the Gopakumar-Vafa (GV) invariants~\cite{Gopakumar:1998ii,Gopakumar:1998jq} describing instanton contributions of branes wrapping the holomorphic curves of a CY acquires direct relevance for the low-energy effective field theories (EFTs) derived from models in string phenomenology.

For us this contact happens for string theory models of axion inflation and dS vacua arising as uplifts of anti-de-Sitter (AdS) vacua, as the non-perturbative quantum corrections encoded by the GV invariants can provide a controlled lifting of flat directions left over in the complex structure (c.s.) moduli space by properly choosing fluxes in type IIB string theory CY orientifold flux compactifications~\cite{Hebecker:2015rya,Hebecker:2020ejb}. The work of~\cite{Hebecker:2015rya,Hebecker:2020ejb} uses the ability to arrange the desired ratios of complex structure moduli VEVs by tuning 3-form fluxes to generate controlled of left-over flat quasi-axion directions in c.s. moduli space near its large-complex-structure point from the instanton contributions encoded by the GV invariants. Moreover,~\cite{Hebecker:2015rya} shows that properly choosing the fluxes can generate flat axion valleys with a large path length on a small fundamental domain, which allows to generate inflationary dynamics once the long flat valley is lifted by the GV-controlled instanton effects.

Based on these literature results, we show in this paper that having a large database of CYs with known GV invariants in hand, we can use CYs with a built-in hierarchy of the lowest-degree GV invariants to collaborate with the tuning of c.s. moduli VEV hierarchies in controlling the instanton contributions to the scalar potential, and in some cases remove the need to tune hierarchical c.s. VEVs. Using the set of complete intersection CYs (CICYs) in projective ambient space as an example database, we provide explicit examples of the required GV invariant hierarchies necessary to alleviate hierarchical arrangements for the c.s. moduli VEVs controlling the relevant instanton contributions, as well as for the control regimes studied in~\cite{Hebecker:2015rya,Hebecker:2020ejb}. Together with the existing results, this provides explicit examples for the mechanism outlined in~\cite{Hebecker:2015rya,Hebecker:2020ejb} to generate both dS vacua and natural-inflation-like slow-roll inflation with the c.s. moduli sector from fluxes and GV invariant controlled quantum corrections alone.

However, it is important to note that the mechanism discussed here can provide a controlled uplift only in a consistent setting with full moduli stabilization. Without reviewing the full discussion of either KKLT-type~\cite{Kachru:2003aw,Demirtas:2019sip,Demirtas:2020ffz} or Large Volume Scenario (LVS)~\cite{Balasubramanian:2005zx} type stabilization of the K\"ahler moduli forming the lightest moduli sector, their respective requirements of either a full set of $h^{1,1}$ rigid 4-cycles or a CY with $h^{1,1}<h^{2,1}$ imply that our explicit examples of this paper show the existence of the ingredients intrinsic to the uplift/inflation sector itself but they cannot be made to work as full examples in particular in the context of LVS. This is because we are able to construct the GV invariants explicitly for the mirror symmetry partner CYs of the CICYs (the so-called mirror CICYs), and these mirror CICYs have $h^{1,1}>h^{2,1}$ owing to the fact that all CICYs themselves have negative Euler characteristic ($\chi=2(\tilde{h}^{1,1}-\tilde{h}^{2,1})<0$).\footnote{Throughout all the paper, we denote the CICYs with a tilde, e.g. $\tilde{X}$ , and $\tilde{h}^{1,1}$ or $\tilde{h}^{2,1}$ respectively $h^{1,1}(\tilde{X})$ and $h^{2,1}(\tilde{X})$. \label{foot:h11h21conv}} 

Rendering our explicit mirror CICY examples of the uplift/inflation mechanism into full examples including moduli stabilization might work upon invoking the KKLT mechanism of non-perturbative K\"ahler moduli stabilization (which does not put a requirement on the sign of $\chi$). However, for the mirror CICYs this would require ``rigidifying" a potentially large number of 4-cycles to remove the potential excess of zero-modes from the required 7-brane stacks providing the non-perturbative quantum corrections responsible for K\"ahler moduli stabilization.

We would like to point out further that these difficulties may be overcome if it can be shown that many of the $\tilde{h}^{1,1}>1$ CICYs and CYs of the Kreuzer-Skarke construction~\cite{Kreuzer:2000xy} set have mirror duals satisfying the Greene-Plesser construction~\cite{Greene:1990ud} properties. For some examples of CICYs in weighted projective ambient spaces (such as, e.g., $\mathbb{P}^4_{11169}$) with $\tilde{h}^{1,1}>1$ this is known to be the case. If such ``Greene-Plesser pairs" can be shown to be ubiquitous, they can potentially be used to transport the GV invariants computation and subsequent construction of the c.s. quasi-axion dS/inflation mechanism done for the mirror CY with small $h^{2,1}<h^{1,1}$ back to the original CY with small $\tilde{h}^{1,1}<\tilde{h}^{2,1}$ such that there is a ``Greene-Plesser" invariant subsector of all c.s. moduli governed by the same triple intersection numbers and GV invariants as the c.s. moduli of the mirror CY. \\
Let us expand on this idea. Crucially, we compactify not on a mirror CICY $X$ but on a CICY $\tilde{X}$ itself, or on any CY with $\tilde{h}^{1,1}<\tilde{h}^{2,1}$. The underlying structure of the Greene-Plesser construction goes as follows. For a Greene-Plesser type CY manifold $\tilde{X}$ with $\tilde{h}^{1,1}<\tilde{h}^{2,1}$ there is an orbifold symmetry $\tilde \Gamma$ by which $\tilde{X}$ can be modded out. Out of the $\tilde{h}^{2,1}$ c.s. moduli of $\tilde{X}$, $\tilde{h}^{2,1}_+$ will be even under the orbifold group action, and $\tilde{h}^{2,1}_-$ will be odd. It turns out that $\tilde{h}^{2,1}_+=\tilde{h}^{1,1}$.
The resulting orbifold $\tilde X/\tilde \Gamma$, after resolving the singularities, has $h^{2,1}(\mathcal{R}(\tilde X/\tilde \Gamma))=\tilde{h}^{2,1}_+$ c.s. moduli and $h^{1,1}(\mathcal{R}(\tilde X/\tilde \Gamma)) = \tilde{h}^{2,1}$  K\"ahler moduli, where $\mathcal{R}$ indicates the resolved manifold. This manifold $\mathcal{R}(\tilde X/\tilde \Gamma)$ therefore has the Hodge diamond of the mirror $X$ and can be shown to be diffeomorphic to $X$. By the structure of the modding action $\tilde\Gamma$ and the resolution procedure one can now show that the periods and prepotential (including instanton corrections) of the $\tilde{h}^{2,1}_+$ c.s. moduli of the CICY are identical to the period and the prepotential (including instanton corrections) for the $h^{2,1}(X)=h^{2,1}(\mathcal{R}(\tilde X/\tilde \Gamma))$ c.s. moduli of $X \cong \mathcal{R}(\tilde X/\tilde \Gamma)$. Finally, the prepotential for the $h^{2,1}$ c.s. moduli of $X$ is written by definition in terms of the GV invariants of its mirror, which is $\tilde{X}$. Hence we conclude that the prepotential of the $\tilde\Gamma$-invariant $\tilde{h}^{1,1}=\tilde{h}^{2,1}_+$-dimensional subsector of the $\tilde{h}^{2,1}$-dimensional c.s. moduli space of $\tilde X$ can be computed from the GV data of $\tilde{X}$ itself.\\
In this case $\chi<0$ examples realizing the uplift/inflation mechanism might be abundant, for which then the most efficient variant of LVS~\cite{Cicoli:2016chb} requiring only a single ED3-instanton correction may solve K\"ahler moduli stabilization in many cases. We leave the extension of our mechanism to non-trivial Greene-Plesser pairs to future work.

Although in specific cases algorithms for the computation of such invariants are known and already implemented~\cite{Hosono:1993qy,Hosono:1994ax}, the computation is most of the time tedious and long. It is then useful to have at hand the list of the GV invariants up to a certain degree for the most used and common CYs used in type II string compactification. We decided to compute the GV invariants for those CICYs with $\tilde{h}^{1,1}\leq 9$ up to degree $10$. We used the algorithm introduced in~\cite{Hosono:1993qy,Hosono:1994ax}, which we review briefly in Appendix~\ref{sec:INSTANTON}. These numbers are accessible from the website \href{https://www.desy.de/~westphal/GV_CICY_webpage/GVInvariants.html}{link} where they are divided by the value of $\tilde{h}^{1,1}$. Each file is named by the number of the CICY using the convention of the list introduced in~\cite{Anderson:2017aux}. In Appendix~\ref{sec:INSTANTON} we give also a pseudocode in Mathematica useful to extract those numbers for practical use.

The ensuing discussion of this paper about the explicit possibility to construct the models we propose by using the hierarchy among the GV invariants is based on the computation of these numbers using \texttt{INSTANTON}~\cite{Klemm:2001aaa} and the scan of our database. 

While computing the GV invariants, we noticed that sometimes the invariants for a pair of  CICYs sharing the same cohomology were the same up to permuting the degrees of the associated curve. This caused us to look for redundancies in the new list of CICYs given by~\cite{Anderson:2017aux}, since, as the authors pointed out, their database contained all necessary inputs to check if Wall's theorem~\cite{WALL1966} could have been applied. We looked for favorable non-product CICYs that can be made equal by a permutation of the basis of $H^4$ of the CICYs. We list all tuples that are equal up to a permutation of the basis of $H^4$ that we found in Appendix~\ref{sec:Tableredundancies} divided by $\tilde{h}^{1,1}$. In that list, some generic transformation involving the CICYs with $(\tilde{h}^{1,1},\tilde{h}^{2,1})=(14,16)$ and $\tilde{h}^{1,1}=15$ are missing, because of long computational time. However, we found more equivalence classes of redundant CICYs with respect to the one found in~\cite{Anderson:2008uw} using the old database list. In Appendix~\ref{sec:CICYreview} we analyze the distribution of the redundant CICYs for different $\tilde{h}^{1,1}$ and related to the total number of favorable CICYs in the database. 

On the \href{https://www.desy.de/~westphal/GV_CICY_webpage/GVInvariants.html}{website}, we link a Mathematica notebook containing the transformation matrices that allow to transform the CICYs belonging to the same equivalence class. We do not provide the matrix for all redundancies, but with the help of Appendix~\ref{sec:Tableredundancies} it is possible to obtain all the other matrices by a simple multiplication of the matrices we provide.

We focused on the redundancies given by permutations of the basis element of $H^4$, so we leave open the possibility that in the CICY list given by~\cite{Anderson:2017aux} there are more redundancies for more generic transformations (in the spirit of what has been done, for instance, in~\cite{He:1990pg}, although for rational cohomology). 

The paper is organized as follows. In Section~\ref{sec:InflationGV} we revisit the inflationary model proposed in~\cite{Hebecker:2015rya} and we propose an alternative model that uses the GV invariants of the lowest degree to avoid creating the hierarchy between the imaginary parts of the c.s. moduli. In Section~\ref{sec:UpliftGV} we consider the uplift model described in~\cite{Hebecker:2020ejb} in LVS and we reinterpret it using the GV invariants for the instantonic corrections of the c.s. moduli involved. In Section~\ref{sec:InflationUpliftGV} we combine the two GV-inspired inflation and uplift models. We discuss the effective inflaton potential that is generated, which is no longer of the pure natural type. We also compute the tunneling transition between an inflationary saddle point and its lower next neighbor. We conclude with a general discussion in Section~\ref{sec:Conclusions}. In Appendix~\ref{sec:CICYreview} we review the general properties of favorable CICYs and we analyze the redundancies we have found in the CICYs list. The list of all redundancies we found in the CICY list divided by $\tilde{h}^{1,1}$ is in Appendix~\ref{sec:Tableredundancies}. 
In Appendix~\ref{sec:INSTANTON} we give a brief description of the algorithm introduced in~\cite{Hosono:1993qy,Hosono:1994ax} for the computation of the GV invariants in the case of a CICY and we explain how to access to the GV invariants in the database on the website \href{https://www.desy.de/~westphal/GV_CICY_webpage/GVInvariants.html}{link}. We also comment on some properties that the GV invariants in the database enjoy.

\section{Winding inflation from Gopakumar-Vafa hierarchies}
\label{sec:InflationGV}

In \cite{Hebecker:2015rya}, the authors present a model of large field inflation for a Calabi-Yau orientifold $X$ compactification of type IIB superstring theory. The inflationary sector and dynamics arise from two c.s. moduli $u, v$.  At the leading order, all other $h^{2,1}(X)-2$ moduli are stabilized at their minimum by fluxes \cite{Giddings:2001yu}. The same happens for the imaginary parts of $u$ and $v$ as well as the real part of a particular linear combination of these two moduli, $\re(Mu+N v)$, with $N\gg M$ ($M$ and $N$ being integer flux numbers).\footnote{We are using a different convention with respect to the one in~\cite{Hebecker:2015rya}, instead we are using the convention of~\cite{Hebecker:2020ejb}. One can restore the convention of~\cite{Hebecker:2015rya} setting $M=1$, $N\longrightarrow -N$ and $z \longrightarrow 2\pi z$.} However, $\re(u)$ is a flat direction in the superpotential, and it is lifted by the exponential terms coming from the instantonic corrections. This term induces a cosine potential for this field which then displays
the effective dynamics of a single slow-rolling axion-like field. Therefore, the field $\re(u)$ is moving along a winding trajectory, whose length is parameterized by the linear combination $\re(Mu+N v)$.

Crucially, the model achieves a long winding trajectory working in a regime of large complex structure for some of the c.s. moduli. Moreover, in order for the winding trajectory to exist, the authors of \cite{Hebecker:2015rya} require that the F-term conditions stabilize $u$ and $v$ such that 
\begin{equation}
e^{-\im(u)}\ll e^{-\im(v)}\ll 1\fstop
\end{equation}
As another assumption, they choose appropriate flux integers so that $u$ and $v$ appear only linearly in the superpotential and include only the instantonic contribution coming from $v$. The authors proceed defining an expansion parameter
\begin{equation}
    \epsilon=e^{-\im(v)}\coma
\end{equation}
and expand the K\"ahler potential and the superpotential at leading order in $\epsilon$. The point is that the F-term conditions stabilize in general all c.s. moduli and the axio-dilaton, but the presence of $\re(Mu+Nv)$ in the superpotential breaks one of the two remaining shift symmetries of $u$ and $v$. The shift symmetry parameterized by $\re(u)$ (which does not appear neither in the K\"ahler potential nor in the superpotential) is a flat direction before introducing the corrections proportional to $\epsilon$. Such corrections generate an oscillating potential, responsible for the inflationary period.\footnote{See also~\cite{Kobayashi:2015aaa} for other realizations based on the mechanism of~\cite{Hebecker:2015rya}.}

We argue that there is another interesting way to realize this hierarchy, by exploiting some properties of the geometry of the extra dimensions. In the following, we develop this idea.

We consider a type IIB Calabi-Yau orientifold $X$ with $h^{2,1}_{-}(X)=2$ c.s. moduli $\{z^{i}\}$ and $h^{1,1}_+$ K\"ahler moduli (also called volume moduli). $X$ has a mirror $\tilde{X}$ with $\tilde{h}^{1,1}_{+}=2$. For the sake of concreteness, we take $\tilde X$ to be a CICY.\footnote{See Appendix \ref{sec:CICYreview} for a brief review of this class of CY manifolds, and \cite{Anderson:2018pui,Hubsch:1992nu} for a more comprehensive treatment.}

We assume further that stabilization of the c.s. moduli, the axio-dilaton and the K\"ahler moduli proceed in hierarchical steps with the following characteristics
\begin{itemize}
\item We supplement the Calabi-Yau orientifold compactification of type IIB string theory by quantized 3-form fluxes as in~\cite{Giddings:2001yu}. These generate a scalar potential for the c.s. moduli and the axio-dilaton, stabilization at mass scales typically somewhat below the KK scale. At this level the K\"ahler moduli remain flat direction in a so-called $4$d ${\cal N}  = 1$ no-scale compactification.
\item Taking into account both perturbative and non-perturbative quantum corrections stabilize the K\"ahler moduli at a lower mass scale in either a supersymmetric (KKLT scenario~\cite{Kachru:2003aw}) or supersymmetry breaking (LVS scenario~\cite{Balasubramanian:2005zx}) AdS vacuum. The stabilization of the K\"ahler moduli must proceed such, that large enough values for the 4-cycle volumes (the individual K\"ahler moduli) as well as the total Calabi-Yau volume are obtained to guarantee decoupling of the volume moduli stabilization from the flux stabilization of the c.s. moduli.
\item A final step of controlled further supersymmetry breaking and uplifting (a positive contribution to vacuum energy) is needed to generate classical stable dS vacuum at the end (see e.g.~\cite{Kachru:2003aw}).
\end{itemize}
We will see that our analysis shows the ingredients of the winding inflation and uplift mechanisms below to be expected to work in any string vacuum satisfying the above generic characteristics.

The K\"ahler potential of the c.s. moduli for the effective $4$d supergravity model in this setup is
\begin{equation}
	\begin{split}
		K=-\ln&\left( -\frac{4}{3} \kappa_{ijk}\im(z^i)\im(z^j)\im(z^k)+i c + \right.\\ &  - 2\sum_{\beta_1,\beta_2}^\infty n_{\beta_1,\beta_2}\left(\text{Li}_3\left(e^{i \beta_i z^i}\right)+\text{Li}_3\left(e^{-i \beta_i \overline{z}^i}\right)\right)+\\
		&\left.-2\sum_{\beta_1,\beta_2}^\infty n_{\beta_1,\beta_2}\beta_i \im(z^i) \left(\text{Li}_2\left(e^{i \beta_i z^i}\right)+\text{Li}_2\left(e^{-i \beta_i \overline{z}^i}\right)\right) \right)\fstop
	\end{split}
	\label{eq:Kahler1}
\end{equation}
Here, $\kappa_{ijk}$ are the triple intersection numbers of $\tilde{X}$, while 
\begin{equation}
	c=-\frac{i}{4\pi}\zeta(3)\chi(X)\coma
\end{equation}
where $\chi(X)$ is the Euler characteristic of the compactification manifold $X$, $\zeta(3)\simeq 1.202$, and $\text{Li}_n(x)$ is the polylogarithm function. The quantities $n_{\beta_1,\beta_2}$  in~\eqref{eq:Kahler1} are the genus $0$ GV invariants, counting the number of holomorphic curves of genus $0$ in a given homology class $[\beta]=[\beta_1,\beta_2]$ of $\tilde{X}$. Such quantities will play a prominent role in our proposal. For reviews see~\cite{Hosono:1994av,Hori:2003ic}.

Now, a few comments about the above K\"ahler potential are in order.
The $4$d ${\cal N}=1$ K\"ahler potential in Eq.~\eqref{eq:Kahler1} is obtained by a projection of the underlying $4$d $\mathcal{N}=2$ K\"ahler potential to the orientifold-even subsector. The underlying $\mathcal{N}=2$ K\"ahler potential itself is obtained in a large complex structure (LCS) limit by mirror symmetry considering the instantonic quantum corrections and it is the same, for instance, of~\cite{Hebecker:2015rya}. Later on in our set up, we stabilize our complex structure moduli at moderate LCS to be slightly larger than $\mathcal{O}(1)$ and never larger than $\mathcal{O}(10)$, so that the use of the LCS limit is justified.

It is now important to note that the truncated K\"ahler potential in Eq.~\eqref{eq:Kahler1} is tree-level with respect to genuine ${\cal N}=1$ quantum corrections. 
For a general $\mathcal{N}=1$ orientifold background, such ${\cal N}=1$ quantum corrections will mix K\"ahler and complex structure moduli space. Indeed, the factorization is only preserved at tree-level~\cite{Grimm:2004uq}. Once quantum corrections are taken into account there are mixing terms that break the factorization. These string loop corrections to the tree-level K\"ahler potential $K_0$ are suppressed inverse powers of the volume of the compactification space (see for instance~\cite{vonGersdorff:2005bf,Berg:2005ja,Berg:2007wt,Cicoli:2008va}). Moreover, they possess a particular structure and scaling property which leads to K\"ahler potential corrections $\delta K\sim {\cal V}^{-p}$ appearing in the scalar potential as $\delta V \sim  {\cal V}^{-2-2p}$ instead of the expected scaling  $\sim  {\cal V}^{-2-p}$ (the factor ${\cal V}^{-2}$ arises from the prefactor $e^{K_0}$ in the scalar potential). This automatic cancellation of the $\sim  {\cal V}^{-2-p}$-terms in the scalar potential is called `extended no-scale'. 

Next, the string loop corrections do depend in their coefficients on the c.s. moduli (e.g. via Eisenstein functions~\cite{Berg:2005ja}). These functions become large for parametrically large values of the c.s. moduli. However, the full loop correction coefficients are also suppressed by the usual $1/(16\pi^2)$ loop factors. 

Hence, as long as the dynamics of winding inflation or winding uplifts is realized using stabilized c.s. values at \emph{moderately} large complex structure (corresponding to c.s. moduli VEVs ${\rm Im}\, z_i\gtrsim {\cal O}(1)$), then the extended no-scale structure of the string loop corrections ensures that, already for quite moderate values of the stabilized volume ${\cal V}$, the induced scalar potential terms are subdominant to any parts of $V$ induced from fluxes, non-perturbative corrections and/or $\alpha'$-corrections used for moduli stabilization and the winding c.s. axion dynamics in e.g. the KKLT or LVS scenarios. 
Hence, provided these conditions are satisfied we can neglect the string loop corrections which would spoil the factorization of the moduli space. 

We can now introduce the usual Gukov-Vafa-Witten superpotential~\cite{Gukov:1999ya}
\begin{equation}
	W=\left(N_F-\tau N_H\right)^T \cdot \Sigma \cdot \Pi\coma
\end{equation}
where $N_F,N_H\in\ZZ$ are flux integers coming from the integration of $F_3$ and $H_3$ on a symplectic base of the 3-cycles of the orientifold CY, $\tau$ is the $10$d axio-dilaton and 
\begin{equation}
	\Sigma=\left(\begin{array}{cc} 0 & -\ID \\ \ID & 0 
	\end{array}\right)\fstop
\end{equation}
$\Pi$ is the period vector with entries
\begin{equation}
	\Pi=\left(\begin{array}{*3{>{\displaystyle}c}p{5cm}}
		1\\
		z^i\\
		\frac{1}{2}\kappa_{ijk}z^jz^k+\frac{1}{2}a_{ij}z^j+b_i-\sum_{\beta_1,\beta_2}^\infty n_{\beta_1,\beta_2}\beta_i\text{Li}_2\left(e^{i \beta_i z^i}\right)\\
		-\frac{1}{3!}\kappa_{ijk}z^iz^jz^k+b_iz^i+\frac{c}{2}+2i\sum_{\beta_1,\beta_2}^\infty n_{\beta_1,\beta_2}\text{Li}_3\left(e^{ i \beta_i z^i}\right)-\sum_{\beta_1,\beta_2}^\infty n_{\beta_1,\beta_2} \beta_i z^i \text{Li}_2\left(e^{ i \beta_i z^i}\right)
	\end{array}\right)\fstop
\end{equation}
Here $a_{ij}$ are related to the triple intersection numbers, while $b_i$ are related to the intersections of the second Chern class and the divisors of $\tilde{X}$.\footnote{Explicit expressions in their convention can be found e.g. in~\cite{Demirtas:2019sip,Demirtas:2020ffz,Blumenhagen:2020ire}.} 

Differently from \cite{Hebecker:2015rya}, in our setup we assume that the F-term conditions stabilize $z_1=u$ and $z_2=v$ in such a way that their imaginary parts are comparable, i.e.
\begin{equation}
	\im(u) \sim \im(v)\fstop
	\label{eq:cond0}
\end{equation}
The requested hierarchy which leads to a winding trajectory is then realized by considering
\begin{equation}
	n_{1,0}e^{-\im(u)}\ll n_{0,1}e^{-\im(v)}\ll 1\coma
\end{equation}
provided that the corresponding GV invariants $n_{0,1}$ and $n_{1,0}$ satisfy 
\begin{equation}
	n_{0,1} \gg n_{1,0}\fstop
	\label{eq:cond1}
\end{equation} 
In order for this hierarchy to be not spoiled by higher instanton effects, we further need to require that
\begin{equation}
	\im(u)\sim \im(v) \gg \ln\frac{n_{0,2}}{n_{0,1}}\coma
	\label{eq:cond2}
\end{equation}
and 
\begin{equation}
	\im(u) \gg \ln \frac{n_{1,1}}{n_{0,1}} \,\,\,\text{ and } \,\,\,\im(v)\gg \ln \frac{n_{1,1}}{n_{1,0}}\fstop
	\label{eq:cond3}
\end{equation}
If \cref{eq:cond2,eq:cond3} are satisfied, all other contributions coming from higher order GV invariants are suppressed by the exponential terms and we can disregard them.

To check if~\cref{eq:cond1,eq:cond2,eq:cond3} can be realized, we scanned the GV invariants of the CICYs with $\tilde{h}^{1,1}=2$, and we found that the hierarchy among the invariants for this inflationary model can be achieved for the CICYs $7819$, $7823$, $7840$, $7867$, $7869$, $7885$, $7886$ and $7888$.\footnote{Modulo redundancies that we discuss in Appendix~\ref{sec:CICYreview} and we list in Appendix~\ref{sec:Tableredundancies}. We are using the numeration of the CICYs as in~\cite{Anderson:2017aux}.} Using these CICYs, the ratio in Eq.~\eqref{eq:cond1} is varying from $31.5$ to $160$. Interestingly, these CICYs also have the invariants $n_{0,m}$  way larger and monotonically increasing with respect to $n_{0,1}$,
and $n_{1,1}$ is equal or a little larger. We need then to fix the expectation values for the imaginary parts of $u$ and $v$ to be larger than the ratio of $n_{1,1}$ and $n_{0,1}$. 

We now need to identify a small $\varepsilon$ parameter, as in~\cite{Hebecker:2015rya}, to get the inflationary potential via a perturbative expansion. The natural definition we adopt is
\begin{equation}
	\varepsilon=n_{0,1}e^{-\im(v)}\fstop
	\label{eq:cond4}
\end{equation} 
Eq.~\eqref{eq:cond4} gives another condition on the values that $\im(v)$ (and $\im(u)$) can assume, since we want $\varepsilon \ll 1$. Notice that requiring $\epsilon \ll 1$ implies that $\im(u)$ and $\im(v)$ are stabilized at large complex structure. In general, this condition alone is sufficient to satisfy all previous ones for the CICYs for which this hierarchy can be realized. 

It is then possible to proceed as in~\cite{Hebecker:2015rya}. At leading order $\im(u)$, $\im(v)$, the axio-dilaton as well as the linear combination $\re(Mu+N v)$ are stabilized at the minimum. The only remaining flat direction is, once again, aligned with $\re(u)$. To proceed with the lifting to get the inflationary potential, we then repeat the discussion already presented in~\cite{Hebecker:2015rya} in more detail. 

It is convenient to reparameterize the fields as
\begin{equation}
	\phi \equiv u \,\text{ and }\,\psi \equiv Mu+Nv \coma
	\label{eq:phipsi}
\end{equation}
and we thus require $N>M$ to have one of the winding directions which is longer than the other.
In this way, the expansion parameter becomes
\begin{equation}
	\varepsilon= n_{0,1}e^{-\im(v)}=n_{0,1}e^{- \frac{\im(\psi)-M \im(\phi)}{N}}\coma
\end{equation}
and 
\begin{equation}
	n_{0,1}e^{ i v}= n_{0,1}e^{ i\frac{\psi-M\phi}{N}}=\varepsilon\, e^{ i \frac{\re(\psi)-M\re(\phi)}{N}}\fstop
\end{equation}
By choosing appropriately the fluxes and introducing the term $W_0(\tau)$ which includes all the fields already stabilized at leading order by the F-terms, we can write the superpotential as 
\begin{equation}
	W= W_0(\tau) +f(\tau)\psi+\varepsilon\,  g_{0,1}(\tau,\psi,\im \phi) \,e^{-i \frac{M \re \phi}{N}}+\mathcal{O}(\varepsilon^2)\coma
\end{equation}
where $g_{0,1}(\tau,\psi,\im \phi)$ is a function of all stabilized fields. We can repeat the same discussion in terms of K\"ahler potential, obtaining
\begin{equation}
	K=K_0(\tau,\psi,\im(\phi))+\varepsilon\, \tilde{g}_{0,1}(\tau,\psi,\im(\phi))\,e^{-i \frac{M \re \phi}{N}}+\mathcal{O}(\varepsilon^2)\fstop
\end{equation}

We have shown that using the hierarchy given by the GV invariants, we could revisit the model introduced in~\cite{Hebecker:2015rya} keeping the expectation values of the c.s. moduli to be at the same order. To conclude this analysis, let us comment on the inflaton potential. The scalar potential for the c.s. moduli sector and the axio-dilaton is given by
\begin{equation}
	V=e^K K^{I \bar{J}} D_I W D_{\bar{J}}\overline{W}\fstop
	\label{eq:VInfgen}
\end{equation}
At zeroth order in $\varepsilon$, $D_I W=0$ sets $ \tau,\,\psi,\,\im(\phi)$ to their minimum and we are left with a flat direction parameterized by $\varphi=\re(\phi)$. 
This flat direction is lifted by the first order corrections in $\varepsilon$ to $K_0$ and $W_0$, which induce a shift in the VEVs of the other moduli. To see this, it is useful to write the structure of the F-terms as
	\begin{equation}
		D_IW=D_I |_0W_0+K_{0,I}\Delta W_{GV}+\Delta K_{GV,I}W_0\equiv D_I W|_0+\Delta D_IW|_{GV}\fstop
	\end{equation}
Since on the supersymmetric flux vacua we have $D_iW|_0=0$, this entails that the scalar potential along $\varphi$ is lifted by the GV corrections at ${\cal O}(\epsilon^2)$, because the non-vanishing potential at the SUSY locus of all other fields is given by
	\begin{equation}
		V_{\text{inf}}\sim e^{K_0}(K^0)^{I\bar J}\Delta D_IW|_{GV}\Delta D_{\bar J}\bar{W}|_{GV}\fstop
		\label{eq:VinfTh}
    \end{equation}
 To give an explicit expression for the effective inflationary axion-like potential in~\eqref{eq:VinfTh}, in~\cite{Hebecker:2015rya} the authors make an orthogonal transformation on Eq.~\eqref{eq:VInfgen} to diagonalize the K\"ahler metric. We can define $\varphi=\re(\phi)$, so that the potential, splitted in real and imaginary parts of the moduli, takes the following form
 \begin{equation}
     V=e^{K_0}\sum_{\alpha=1}^{6}\tilde{w}_\alpha^2\coma
     \label{eq:VInflw}
 \end{equation}
 where
 \begin{equation}
     \tilde{w}_\alpha=\tilde{a}_\alpha+\epsilon\left[\tilde{b}_\alpha \cos\left(\frac{M\varphi}{N}\right)+\tilde{c}_\alpha \sin\left(\frac{M\varphi}{N}\right)\right]
 \end{equation}
 with $\tilde{a}_\alpha$, $\tilde{b}_\alpha$ and $\tilde{c}_\alpha$ being functions of all moduli. From the classical F-terms, $\tilde{a}_\alpha=0$ for all values of $\alpha$. However, considering the $\mathcal{O}(\epsilon)$ corrections coming from the GV invariants, the VEVs of $\tilde{a}_\alpha$, $\tilde{b}_\alpha$ and $\tilde{c}_\alpha$ get shifted. Since Eq.~\eqref{eq:VInflw} is proportional to $\tilde{w}^2_\alpha$ and we are interested in a potential up to order $\mathcal{O}(\epsilon^2)$, it is sufficient to consider order $1$ corrections in $\epsilon$ only for $\tilde{a}_\alpha$, while keeping at leading order $\tilde{b}_\alpha$ and $\tilde{c}_\alpha$. A further rotation and a change of basis in the fields~\cite{Hebecker:2015rya} cancel all six terms but one combination, which is the inflaton potential
\begin{equation}
	V_{\text{inf}}\left(\varphi\right)\sim e^{K_0} \kappa\, \varepsilon^2 \left[\sin\left(\frac{M}{N}\varphi+\theta\right)\right]^2\sim  e^{K_0} \kappa\, \varepsilon^2  \left[1-\cos\left(2\frac{M}{N}\varphi+2\theta\right)\right] \coma
	\label{eq:Vinfonefield}
\end{equation}
where $\kappa$ encodes numerical and $\tau$-independent factors and $\theta$ is a phase. \\
We shall now pause shortly to comment about the interplay among K\"ahler moduli stabilization and the c.s. axion winding potential. Stabilizing the volume moduli e.g. as prescribed in the KKLT or LVS scenarios leads to a mass hierarchy between the c.s. moduli and K\"ahler moduli (which is more pronounced for the LVS) as well as a hierarchy between the terms of the flux scalar potential fixing the c.s. moduli ${\cal O}\left({\cal V}^{-2}\right)$ and the volume moduli scalar potential (${\cal O}\left(|W_0| {\cal V}^{-2}\right)$ for KKLT, and ${\cal O}\left({\cal V}^{-3}\right)$ for LVS, see the discussion in~\cite{Balasubramanian:2005zx}). Next, the terms of the c.s. axion winding scalar potential are controlled by the GV invariants and the VEVs of the c.s. moduli. However, these VEVs were determined by the $3$-form flux scalar potential and hence receive only suppressed corrections from the stress-energy sources driving K\"ahler moduli stabilization by virtue of the above hierarchies. We conclude that the K\"ahler moduli stabilizing part of the moduli scalar potential, which indeed does in general spoil factorization of the moduli space, will not affect the c.s. moduli stabilization generated c.s. axion winding potential at leading order.

The generalization of the previous discussion to an arbitrary number of c.s. moduli could in principle be straightforward. Consider a Calabi-Yau $X$ with $h^{2,1}>2$ which is mirror to a CICY $\tilde{X}$ and assume that the imaginary parts of all moduli are comparable. Then, consider two different GV invariants $n_{i_1,\ldots i_{\tilde{h}^{1,1}}}$ and $n_{j_1,\ldots j_{\tilde{h}^{1,1}}}$ both of degree $1$. We request that $n_{i_1,\ldots i_{\tilde{h}^{1,1}}}\ll n_{j_1,\ldots j_{\tilde{h}^{1,1}}}$, and furthermore all the other degree $1$ invariants are smaller than those two.

However, by looking at the scanned GV invariants of all CICYs, it is quite hard to find such a hierarchy. Instead, the values of the GV invariants are always more comparable when $\tilde{h}^{1,1}$ of the CICY increases. Therefore, our proposal for a generalization must include a fine-tuning of the VEVs of all moduli but two. We tune the fluxes in such a way that all c.s. moduli are stabilized except for two of them. These two moduli must then be associated to GV invariants which display the hierarchy~\eqref{eq:cond1}. If the imaginary parts of these moduli guarantee that Eq.~\eqref{eq:cond4} is smaller than $1$, we can thus reproduce the procedure above for any CY which is mirror to a CICY with an arbitrary $\tilde{h}^{1,1}$. In particular, we checked in our database that $77$ CICYs display a hierarchy of $1:30$ for two GV invariants, i.e., the same hierarchy we required in this section. Such hierarchy involves the smallest positive GV invariant and the largest one. However, if we relax this requirement and demand a smaller hierarchy, for instance $1:10$, we have that around $23\%$ of all the CICYs can provide such scenario. Notice that all these numbers must be intended as ``at least", as our scan covers the cases up to $\tilde{h}^{1,1}=9$ only.\footnote{The maximum value of the degree GV invariant for the CICYs from $\tilde{h}^{1,1}=1$ to $\tilde{h}^{1,1}=9$ is decreasing with $\tilde{h}^{1,1}$, going from $2875$ of the quintic, i.e. $7890$, to $30$ of the CICYs $1121$, $1127$, $1157$, $1247$, $1258$. It is always more difficult to find the hierarchy we are looking for when you increase $\tilde{h}^{1,1}$. We comment on the properties we found on GV invariants in the database we constructed in Appendix~\ref{sec:INSTANTON}.\label{foot:h11decrease}}

\section{Uplift mechanisms from Gopakumar-Vafa hierarchies}
\label{sec:UpliftGV}

Our next goal is to exhibit the role of the GV invariants when their associated instanton contributions are used to construct a de Sitter uplift.
This was recently done in~\cite{Hebecker:2020ejb} in the context of type IIB Calabi-Yau orientifold compactification in the large complex structure limit. 
By tuning the flux quanta, the authors were able to generate an oscillating potential for the c.s. moduli, involving several cosines. This potential has a sequence of minima of increasing positive vacuum energy contribution, which are responsible for the controlled SUSY breaking. Choosing the parameters of this potential such that the difference between two adjacent minima is smaller than the depth of the scalar potential produced by the stabilization of the K\"ahler moduli, for instance, in LVS~\cite{Balasubramanian:2005zx}, it is possible to realize an uplift of either a KKLT-type or LVS-type AdS vacuum to a de Sitter vacuum.
In their paper, the authors considered both LVS and KKLT~\cite{Kachru:2003aw} setups, as well as type IIA compactifications with fluxes~\cite{DeWolfe:2005uu,Marchesano:2019hfb,Junghans:2020acz,Marchesano:2020qvg}. Here we will focus only on LVS-type vacua.

The situation here is different from what we described in Section~\ref{sec:InflationGV}. In the current case, the authors of~\cite{Hebecker:2020ejb} tune the saxion VEVs $\im(u)$ and $\im(v)$ to be comparable, such that
\begin{equation}
    \epsilon\equiv e^{-\im(u)}\sim e^{-\im(v)}\ll 1\coma
    \label{eq:epsHebUp}
\end{equation}
and the relative magnitude is encoded in the parameter
\begin{equation}
    \alpha \propto e^{\im(v)-\im(u)} \sim \mathcal{O}(1)\fstop
    \label{eq:alphaHebUp}
\end{equation}
By making an analogous discussion as the one performed around Eq.~\eqref{eq:VinfTh}, but with the above assumptions, the resulting potential coming from the F-terms of the superpotential is found~\cite{Hebecker:2020ejb} to be parameterized by\footnote{Notice that here we took $K_0\sim \ln\left(g_s \mathcal{V}^{-2}\right)$ already.}
\begin{equation}
    V(u)=\frac{g_s}{\mathcal{V}^2}\kappa\,\epsilon^2\left[\cos\left(\re(u)\right)-\alpha\cos\left(\frac{P}{Q}\re(u)\right)\right]^2\coma
    \label{eq:VUplHeb}
\end{equation}
where $\kappa$ contains all the information coming from the K\"ahler metric and the K\"ahler potential, $P$ and $Q$ are flux integers such that $P/Q>1$ and $\mathcal{V}$ is the volume of the CY. This potential has a stationary point when $\re(u)=0$ but, differently from the inflationary potential of Section~\ref{sec:InflationGV}, the value of the minimum is different from zero. Instead we have
\begin{equation}
    V(0)=\frac{g_s}{\mathcal{V}^2}\kappa\, \epsilon^2(1-\alpha)^2\fstop
\end{equation}
From LVS, the supersymmetric minimum of the potential is negative, i.e.
\begin{equation}
    V_{\AdS}=-\mathcal{O}(1)\frac{g_s|W_0|^2\sqrt{\ln\left(\mathcal{V}\right)}}{\mathcal{V}^3}<0\coma
    \label{eq:VLVSAds}
\end{equation}
with $|W_0|$ coming from the stabilization of all c.s. moduli and the axio-dilaton. It is possible to consider the superposition of the LVS potential with~\eqref{eq:VUplHeb} and tune the parameters to get a controlled SUSY breaking and an uplift from AdS to Minkowski or dS vacuum.

The purpose of this section is to argue that it is possible to realize the uplift using a GV hierarchy, to recover a setup similar to the one in~\cite{Hebecker:2020ejb}.
As an example, we will consider type IIB orientifold on $X$, where $X$ is the mirror of a given CICY $\tilde{X}$ with $\tilde{h}^{1,1}=2$. \\
Requiring the VEVs of the saxions $\im(v)$ and $\im(u)$ to be comparable, as in~\cite{Hebecker:2020ejb}, i.e.
\begin{equation}
	\im(u)\sim \im(v) 
	\label{eq:imvimuupl}
\end{equation}
one should look for a CICY $\tilde{X}$ whose degree $1$ GV invariants satisfy
\begin{equation}
	\frac{n_{0,1}}{n_{1,0}}\sim \mathcal{O}(1)\fstop
	\label{eq:upliftcond0}
\end{equation}
In particular, we checked that such CYs exist inside the CICY database. There are $5$ CICYs ($7644$, $7761$, $7799$, $7863$ and $7884$) that have this ratio exactly equal to $1$. Moreover, there are other $17$ CICYs\footnote{Up to redundancies listed in Appendix~\ref{sec:Tableredundancies}.} that have a ratio $\mathcal{O}(1)$ ($7643$, $7668$, $7725$, $7726$, $7758$, $7759$, $7807$, $7809$, $7816$, $7821$, $7822$, $7833$, $7844$, $7853$, $7868$, $7882$ and $7883$).
Imposing Eq.~\eqref{eq:upliftcond0}, we can modify the definition of $\epsilon$ in Eq.~\eqref{eq:epsHebUp} in this context to
\begin{equation}
    \epsilon=n_{0,1}e^{-\im(v)}\sim n_{1,0}e^{-\im(u)}\coma
    \label{eq:equalcond}
\end{equation}
leaving the relative magnitude $\alpha$ as defined in Eq.~\eqref{eq:alphaHebUp}. We have then realized the same setup described in~\cite{Hebecker:2020ejb} with a slightly different definition of $\epsilon$ that keeps into account the values of the GV invariants. Since $n_{0,1}\sim n_{1,0}$, there are no substantial differences with~\cite{Hebecker:2020ejb}, because we have not required a hierarchy either between the VEVs of the axions or among the relevant GV invariants.
However, we have seen in Section~\ref{sec:InflationGV} that there are many CICYs that have GV invariants satisfying Eq~\eqref{eq:cond1}. For those CICYs it is not possible to define $\epsilon$ as in Eq.~\eqref{eq:upliftcond0} keeping the ratio of the VEVs of the saxions $\mathcal{O}(1)$. 
By looking at the CICYs with $\tilde{h}^{1,1}=2$, we see that $6$ CICYs ($7806$, $7808$, $7817$, $7858$, $7873$ and $7887$) have a ratio of GV invariants that will not be able to reproduce the model of~\cite{Hebecker:2020ejb}, if we insist in Eq.~\eqref{eq:imvimuupl}.
This is why we would like to propose another possibility that generates the setup of~\cite{Hebecker:2020ejb}. We could play the same trick we did in Section~\ref{sec:InflationGV}, redefining the parameter $\varepsilon$ of the expansion as in Eq.~\eqref{eq:equalcond}, but compensating for the large ratio between the GV invariants with a specific tuning of the VEVs of the saxions. It is then possible to revisit the model introduced in~\cite{Hebecker:2020ejb} by choosing $\epsilon\ll 1$ in Eq.~\eqref{eq:equalcond} but requiring
\begin{equation}
	\im(v)-\im(u) \sim \ln\left(\frac{n_{0,1}}{n_{1,0}}\right)\fstop
	\label{eq:upliftcond1}
\end{equation}
The relative magnitude in Eq.~\eqref{eq:alphaHebUp} is then modified as
\begin{equation}
    \alpha \propto \frac{n_{1,0}}{n_{0,1}}e^{\im(v)-\im(u)}
    \label{eq:alphaUp}
\end{equation}
and it is still $\mathcal{O}(1)$ due to the condition in~\eqref{eq:upliftcond1}.

Given the definition of $\varepsilon$ in Eq.~\eqref{eq:equalcond}, we can proceed as in the previous section by parameterizing $u$ and $v$ as in~\eqref{eq:phipsi}, i.e. $\psi\equiv P u + Q v$. At leading order in $\varepsilon$, the fields $\tau$, $\psi$ and $\im \phi$ are stabilized in their minimum while $\re \phi$ is left as a flat direction. To uplift this direction, we must consider the first order in the $\varepsilon$ expansion for the superpotential and K\"ahler potential, which read
\begin{equation}
    \begin{split}
    W&= W_0(\tau,\psi) +\varepsilon\,  \left[g_{0,1}(\tau,\psi,\im \phi) \,e^{-i \frac{P }{Q}\re \phi}+h_{0,1}(\tau,\im \phi) \,e^{i \re \phi}\right]+\mathcal{O}(\varepsilon^2)\coma\\
    K&=K_0(\tau,\psi,\im \phi)+\varepsilon\, \left[\tilde{g}_{0,1}(\tau,\psi,\im \phi)\,e^{- i \frac{P }{Q}\re \phi}+\tilde{h}_{0,1}(\tau,\im \phi) \,e^{i \re \phi}\right]+\mathcal{O}(\varepsilon^2)\coma
    \end{split}
    \label{eq:WKUpl}
\end{equation}
where the presence of two contributions in $\varepsilon$ now comes from the requirement in~\eqref{eq:equalcond}.\\
Once again, at the zeroth order in $\epsilon$, $D_IW=0$ sets $\tau$, $\psi$ and $\im(\phi)$ to their minimum and we are left with a flat direction given by $\varphi\equiv\re(\phi)$. The flat direction is lifted by the first order corrections in $\epsilon$ as shown in Section~\ref{sec:InflationGV}. Keeping into account that the superpotential and the K\"ahler potential this time are given by~\eqref{eq:WKUpl}, and performing an analogous rotation of c.s. moduli in~\cite{Hebecker:2020ejb}, the authors suggest a potential of the following form:
\begin{equation}
	V_{\dS}(\varphi)=e^{K_0} \kappa\, \varepsilon^2\left[\cos\left(\varphi+\theta_1\right)-\alpha\cos\left(\frac{P}{Q}\varphi+\theta_2\right)\right]^2\fstop
	\label{eq:Vuplift}
\end{equation}
Here, $\kappa$ encodes numerical and $\tau$-independent factors, $\theta_{1,2}$ are phases and $\alpha$ is the $\mathcal{O}(1)$ parameter introduced in Eq.~\eqref{eq:alphaUp}.

By tuning the phases to zero, the potential has a stationary point at
\begin{equation}
\begin{split}
    V_{\dS}(0)&=e^{K_0} \kappa \,\varepsilon^2(1-\alpha)^2\coma\\
    V_{\dS}''(0)&=2e^{K_0} \kappa\,\varepsilon^2(1-\alpha)\left(\frac{P^2}{Q^2}\alpha-1\right)\coma
\end{split}
\end{equation}
which is a minimum for $Q^2/P^2<\alpha <1$, provided that $P/Q>1$. In~\cite{Hebecker:2020ejb}, then, the authors assume that the potential is given by the sum of~\eqref{eq:VLVSAds} and~\eqref{eq:Vuplift}, i.e.
\begin{equation}
    V(\mathcal{V},\varphi)=V_{\text{LVS}}(\mathcal{V})+\frac{g_s}{\mathcal{V}^2} \kappa\, \varepsilon^2\left[\cos\left(\varphi+\theta_1\right)-\alpha\cos\left(\frac{P}{Q}\varphi+\theta_2\right)\right]^2\fstop
\end{equation}
Finally, it is possible to scan the flux landscape and to tune $\alpha$ to make an uplift from the AdS vacuum to a dS one, imposing at the stationary point the relation for~\eqref{eq:Vuplift}
\begin{equation}
    \kappa\, \epsilon^2 (1-\alpha)^2=\mathcal{O}(1)\frac{|W_0|^2\sqrt{\ln\left(\mathcal{V}\right)}}{\mathcal{V}}\fstop
\end{equation}\\

 We can now discuss a possible generalization of this treatment to $\tilde{h}^{1,1}>2$, analogously to what we did for the inflationary setup in Section~\ref{sec:InflationGV}. Whenever the GV invariants involved in the potential are of the same order of magnitude, it is possible to follow~\cite{Hebecker:2020ejb} again. We provide here an estimate of how many CICYs with $\tilde{h}^{1,1}>2$ could display such feature. In this case we require two CICYs to have at least some degree $1$ GV invariants equal, and that these invariants should be numerically the smallest ones. We require this to be able to stabilize all other c.s. moduli via larger GV invariants and an appropriate tuning of the VEVs. It turns out that at least the $24\%$ of the CICYs\footnote{We stress again that we are looking at the scan we have done, that contains CICYs up to $\tilde{h}^{1,1}=9$.} fulfill such requirement, since they have the smallest positive degree $1$ GV invariant repeated exactly twice on different directions inside the Mori cone. As an example, for the CICYs $7236$ and $6968$ (whose $\tilde{h}^{1,1}=5$) two degree $1$ GV invariants vanish, and two others are equal to $3$. The remaining degree $1$ invariant is equal to $144$ in one case and $117$ in the other. For these CICYs, the hierarchy between the GV invariants is already good enough to realize the uplift described above, provided that we choose the VEVs of the moduli associated to the invariants equal to $3$ in such a way that~\eqref{eq:equalcond} is satisfied.

However, it can also happen that the smallest positive degree $1$ GV invariants are repeated more than twice in different directions inside the Mori cone. In this case, we could fix the VEVs via an appropriate tuning for all c.s. moduli except for two of them, and then choose the imaginary parts of the latter two in such a way that we can realize~\eqref{eq:equalcond} by varying the VEVs of the moduli. 
Such examples will also display the right structure of GV invariants to realize the above discussed uplift mechanism. The percentage of CICYs satisfying these conditions is over $47\%$. Therefore, we conjecture that the uplift mechanism of~\cite{Hebecker:2020ejb}, realized by a GV hierarchy, can be a quite generic construction. 

An important comment is now due. In this section we argue that it is possible to realize the contribution to the uplift coming from the complex structure potential in LVS by choosing to compactify on a CY $X$ whose mirror CY $\tilde{X}$ has a suitable set of GV invariants, and we show that such CY $\tilde{X}$ exists in the CICY database.

However, we stress that one cannot make a realistic complete model for the uplift with this mechanism by compactifying on $X$, because a crucial point for the LVS moduli stabilization to hold is to have the Euler characteristic $\chi< 0$ in the BBHL correction term~\cite{Becker:2002nn} to the K\"ahler potential. All the CICYs have a non-positive Euler characteristic which in turn means that their mirrors have $\chi\geq 0$.

We hope that it is possible to find CYs whose mirrors have the right pattern of GV invariants and at the same time the right sign of $\chi$, and we leave this to further investigation. Perhaps, it is possible to find such examples among the much larger database of CYs realized as the anticanonical hypersurface in a $4$-dimensional toric ambient space~\cite{Kreuzer:2000xy}. Since this database is closed (by construction) under mirror symmetry, half of the CYs there have the right sign for the Euler characteristic. The question is then repeating a scan similar to the one we performed in this paper, to look for right Gopakumar-Vafa structures. We leave this to future work.

\section{Combining GV-inspired inflation and uplifts}
\label{sec:InflationUpliftGV}

In the previous sections we discussed setups where using CYs with hierarchical lowest-degree GV invariants leads to a scalar potential which in the presence of full moduli stabilization can realize winding inflationary models and or a dS uplift of an AdS LVS vacuum very similar and much along the lines of~\cite{Hebecker:2015rya,Hebecker:2020ejb}.

In this section, we ask if it is possible to combine an inflationary sector with an uplift sector, both arising from similar effects as discussed before. The idea is to generalize the examples presented before to a case in which you have more c.s. moduli. To simplify the example, we choose a manifold $X$ whose mirror is a CICY with $\tilde{h}^{1,1}=4$. Let us call the complex structure moduli $u_1$, $v_1$, $u_2$ and $v_2$. At the minimum, their imaginary parts, the axio-dilaton, $\re(M u_1+N v_1)$ and $\re(P u_2+Q v_2)$ are stabilized, but $\re(u_1)$ and $\re(u_2)$ are flat directions when we do not consider the exponential terms. By tuning the fluxes, we can choose
\begin{equation}
	\phi_1=u_1 \coma \psi_1=M u_1+N v_1
\end{equation}
and define the expansion parameter 
\begin{equation}
	\varepsilon_1=n_{0,1,0,0}e^{- \im(v_1)}=n_{0,1,0,0}e^{- \frac{\im(\psi_1)-M\im(\phi_1)}{N}}\fstop
	\label{eq:ep1Infl}
\end{equation}
This definition should remind of the discussion in Section~\ref{sec:InflationGV} for $N>M$ where again the hierarchy among the GV invariants must be
\begin{equation}
	n_{1,0,0,0}e^{-\im(u_1)}\ll n_{0,1,0,0}e^{- \im(v_1)}\fstop
	\label{eq:ep1hierarchy}
\end{equation}
Therefore, we can neglect the contributions coming from the instantonic corrections for $u_1$. The idea is once again to generate an inflationary potential provided that~\eqref{eq:ep1Infl} is smaller than $1$.\footnote{We are choosing the GV invariants associated to $v_1$ arbitrarily, we are not referring to a specific CICY at the moment. We will comment later about the hierarchy that you need among the GV invariants and the VEVs of the moduli.} 

A similar discussion can be carried out for the other two moduli $u_2$ and $v_2$, by introducing
\begin{equation}
	\phi_2=u_2 \coma \psi_2=P u_2+Q v_2
\end{equation}
and
\begin{equation}
	\varepsilon_2=n_{0,0,1,0}e^{- \im(v_2)}=n_{0,0,1,0}e^{- \frac{\im(\psi_2)-P\,\im(\phi_2)}{Q}}\sim n_{0,0,0,1}e^{- \im(u_2)}=n_{0,0,0,1}e^{- \im(\phi_2)}\fstop
	\label{eq:ep2def}
\end{equation}
This time, the instantonic contributions coming from both the moduli $u_2$ and $v_2$ are comparable and must be both kept in the expansion. Such conditions can be obtained by tuning the expectation values and fluxes as in Section~\ref{sec:UpliftGV}, but it is possible to scan over all GV invariants for the CICYs at $\tilde{h}^{1,1}=4$ to check if the hierarchy of Eq.~\eqref{eq:ep1hierarchy} and the condition in Eq.~\eqref{eq:ep2def} can be realized.

Another condition that must be guaranteed is the one controlling the order in which inflation and uplift must happen. What we want to ask is that $\varepsilon_1$ is controlling the dynamics of the inflationary regime at an energy smaller than the one used for the uplift controlled by $\varepsilon_2$. Crucially, we should require that $\varepsilon_1\ll\varepsilon_2$. Since we also want the two regimes to happen (almost) independently, we can assume that the effects of the two expansions are just a superposition of the single effects. The superpotential and the K\"ahler potential after these reparametrizations are
\begin{equation}
	\begin{split}
		W= &\,W_0(\tau,\psi_1,\psi_2)+\varepsilon_1\, g_{0,1,0,0}(\tau,\psi_1,\im \phi_1)e^{-i \frac{M}{N}\re(\phi_1)}+\\
		&+\varepsilon_2\left[g_{0,0,1,0}(\tau,\psi_2,\im \phi_2)e^{-i  \frac{P}{Q}\re(\phi_2)}+h_{0,0,1,0}(\tau,\im \phi_2)e^{i\re(\phi_2)}\right]+\mathcal{O}(\epsilon^2)\coma\\
		K=&\,K_0(\tau,\psi_1,\psi_2,\im \phi_1,\im \phi_2)+\varepsilon_1\, \tilde{g}_{0,1,0,0}(\tau,\psi_1,\im \phi_1)e^{-i \frac{M}{N}\re(\phi_1)}+\\
		&+\varepsilon_2\left[\tilde{g}_{0,0,1,0}(\tau,\psi_2,\im \phi_2)e^{-i  \frac{P}{Q}\re(\phi_2)}+\tilde{h}_{0,0,1,0}(\tau,\im \phi_2)e^{i\re(\phi_2)}\right]+\mathcal{O}(\epsilon^2)\fstop
	\end{split}
\end{equation}
In the previous equations we are neglecting all terms of order $\varepsilon_1^2$, $\varepsilon_2^2$ and $\varepsilon_1\varepsilon_2$. Let us spend some more words about this approximation. Suppose we want to realize the situation described in~\cite{Hebecker:2020ejb} and reviewed in our set-up in Section~\ref{sec:UpliftGV}. The potential is found after having integrated out the other c.s. moduli. 
Similar to the discussion above, the F-terms split as  
	\begin{align}
		D_IW&=D_I |_0W_0+K_{0,I}\Delta W_{GV}^{(\phi_1)}+\Delta K_{GV,I}^{(\phi_1)}W_0+K_{0,I}\Delta W_{GV}^{(\phi_2)}+\Delta K_{GV,I}^{(\phi_2)}W_0\nonumber\\
		&\equiv  D_I W|_0+\Delta D_IW|_{GV}^{(\phi_1)}+\Delta D_IW|_{GV}^{(\phi_2)}\fstop
	\end{align}
	Hence, the total scalar potential at ${\cal O}(\epsilon^2)$ scales as
	\begin{equation}
		V_{\text{tot}}\sim e^{K_0}(K^0)^{I\bar J} \left(\Delta D_IW|_{GV}^{(\phi_1)}+\Delta D_IW|_{GV}^{(\phi_2)}\right) \left(\Delta D_{\bar J}\overline{W}|_{GV}^{(\phi_1)}+\Delta D_{\bar J}\overline{W}|_{GV}^{(\phi_2)}\right)\fstop
	\end{equation}
	This scalar potential has three pieces
	\begin{equation}
		V_{\text{tot}}\sim V_{\text{inf}}^{{\cal O}(\epsilon_1^2)}+V_{\dS}^{{\cal O}(\epsilon_2^2)}+\left.\sqrt{V_{\text{inf}}}\sqrt{V_{\dS}}\right|^{{\cal O}(\epsilon_1\epsilon_2)}\coma
		\label{eq:PotTot}
	\end{equation}
	where $V_{\text{inf}}^{{\cal O}(\epsilon_1^2)}$ and $V_{\dS}^{{\cal O}(\epsilon_2^2)}$ read
\begin{equation}
V_{\text{inf}}(\varphi_1)=e^{K_0}\kappa\, \varepsilon^2_1\left[\sin\left(\frac{M}{N}\varphi_1+\theta_1\right)\right]^2\coma
\end{equation}
\begin{equation}
V_{\text{dS}}(\varphi_2)=e^{K_0}\kappa\, \varepsilon^2_2\left[\cos\left(\varphi_2+\theta_{2,1}\right)-\alpha_2\cos\left(\frac{P}{Q}\varphi_2+\theta_{2,2}\right)\right]^2\coma
\end{equation}
and we have defined $\varphi_1\equiv\re(\phi_1)$ and $\varphi_2\equiv\re(\phi_2)$.
It is easy to see from Eq.~\eqref{eq:PotTot} that $\left.\sqrt{V_{\text{inf}}}\sqrt{V_{\dS}}\right|^{{\cal O}(\epsilon_1\epsilon_2)}$ has the same stationary points with respect to $\varphi_2$ of $V_{\dS}^{{\cal O}(\epsilon_2^2)}$. The hierarchy $\epsilon_1\ll\epsilon_2$ may thus enable us to stabilize into dS using $V_{\dS}^{{\cal O}(\epsilon_2^2)}$ while having a slow-roll inflation valley given by the suppressed cross-term $\left.\sqrt{V_{\text{inf}}}\sqrt{V_{\dS}}\right|^{{\cal O}(\epsilon_1\epsilon_2)}$ modulated by the far stronger suppressed term $V_{\text{inf}}^{{\cal O}(\epsilon_1^2)}$. 

\begin{figure}[htp]
	\centering
	\begin{tikzpicture}
	\node (img) {\includegraphics[width=0.9\textwidth,keepaspectratio]{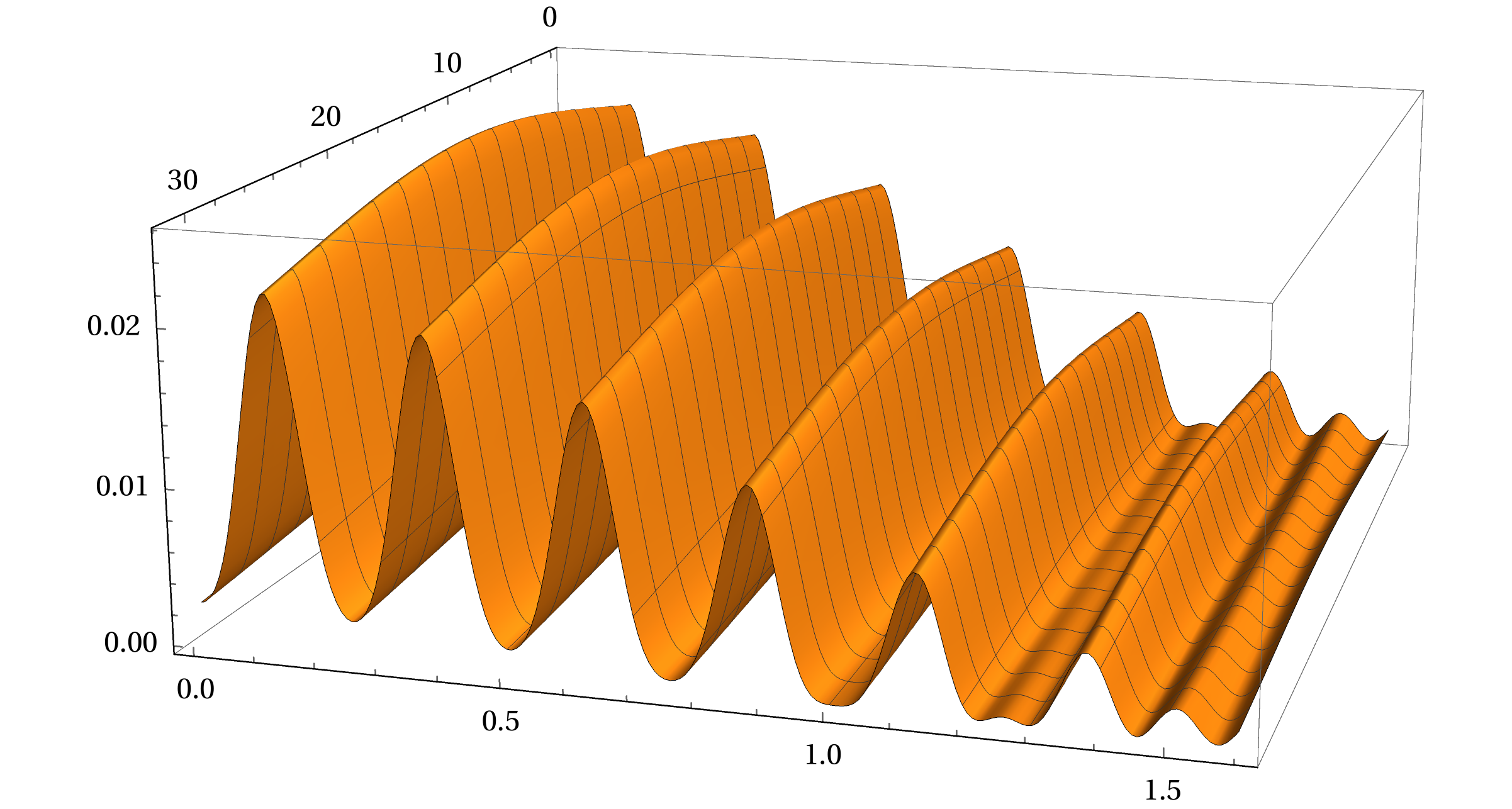}};
	\node[below= of img, node distance=0cm,yshift=1.5cm,xshift=-1cm] {$\varphi_2$};
	\node[above= of img, node distance=0cm,xshift=-3.7cm,yshift=-1.7cm] {$\varphi_1$};
	\node[left=of img, node distance=0cm,rotate=90,anchor=center,yshift=-1.2cm,xshift=-0.5cm] {$V_{\text{tot}}(\varphi_1,\varphi_2)$};
	\end{tikzpicture}
	\caption{An example of the potential in Eq.~\eqref{eq:PotTot}. We use $M/N=1/10$, $P/Q=25$, all the phases zero, $\alpha_2=1$, $\epsilon_1=0.02$, $\epsilon_2=0.1$.}
	\label{fig:PotTot}
\end{figure}
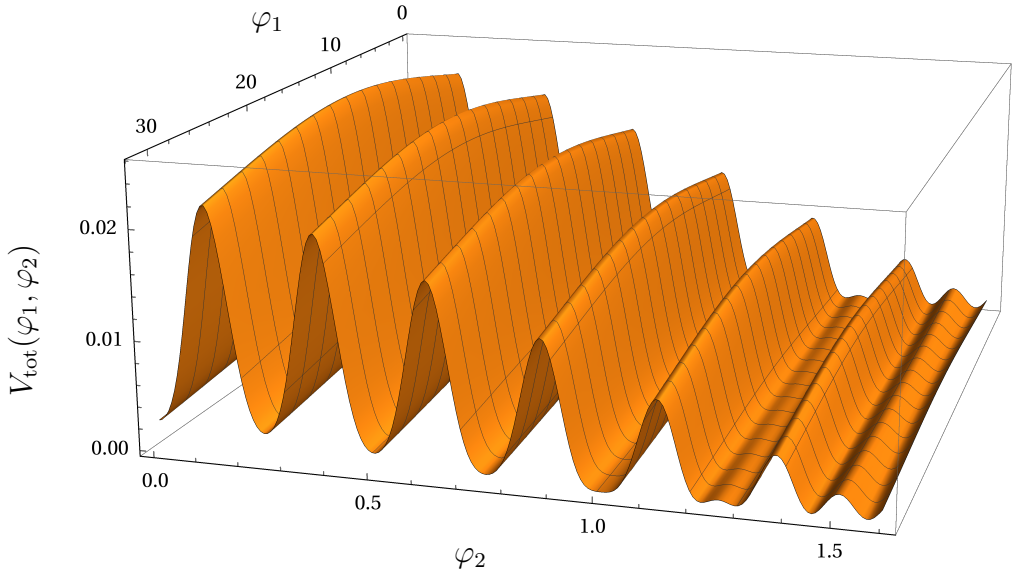

Very interestingly, the effective inflaton potential is no longer of the pure natural inflation type. For instance, a Fourier decomposition of the effective scalar potential 
$V_{\text{eff.}}^{\text{valley}}(\varphi_1)$
in a $\varphi_1$-valley defined by the condition $(\partial_{\varphi_2} V)(\varphi_1)=0$ will generically have the form 
\begin{equation}
V_{\text{eff.}}^{\text{valley}}(\varphi_1)\sim \left[1-\cos\left(2\frac{M}{N}\varphi_1+2\theta_1\right)\right]+\sum\limits_{n\geq 2} c_n\cos(\omega_n\varphi_1)\coma
\end{equation}
with rapidly decreasing $c_n$, frequencies $\omega_n$ being multiples of $2 M/N$. Therefore, we expect the predictions for CMB observables like the spectral tilt $n_s$ and the tensor-to-scalar ratio $r$ to deviate from pure natural inflation. We leave an analysis of the ensuing phenomenology for future work.

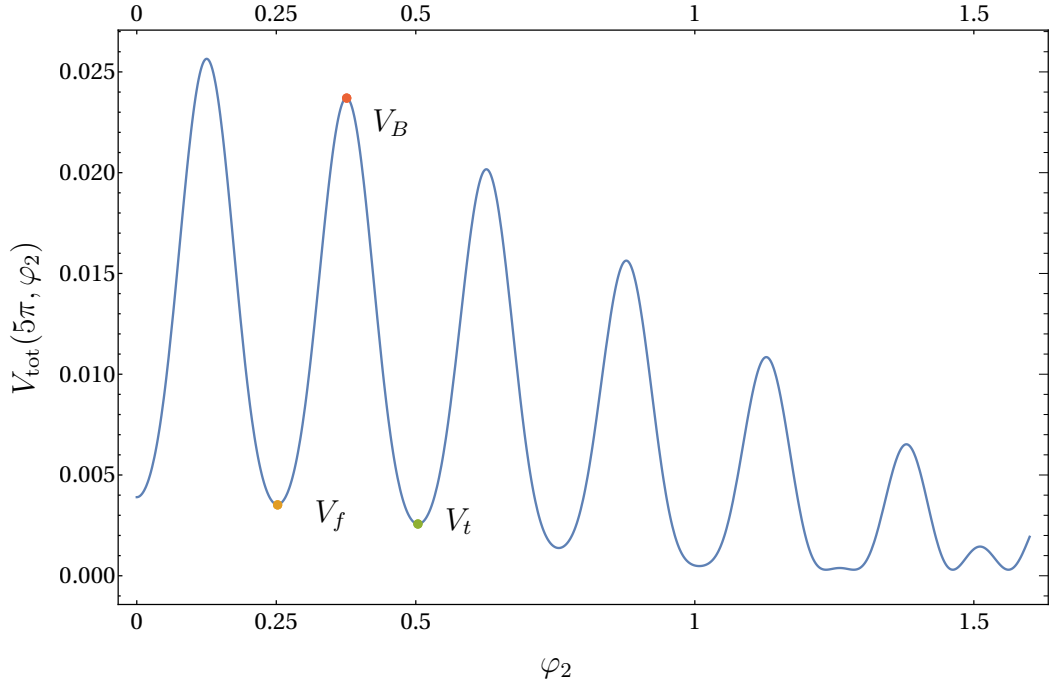
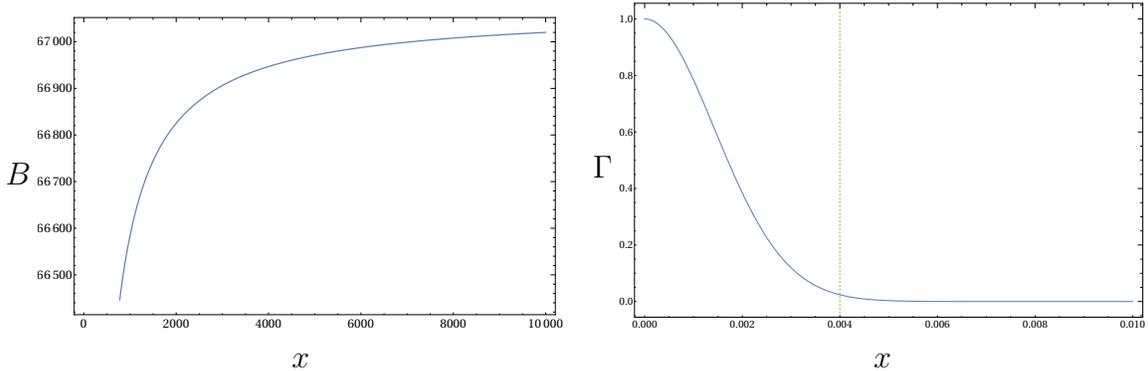
\begin{figure}[htp]
	\centering
	\begin{subfigure}[t]{\textwidth}
		\centering
		\begin{tikzpicture}
		\node (img) {\includegraphics[width=0.85\textwidth,keepaspectratio]{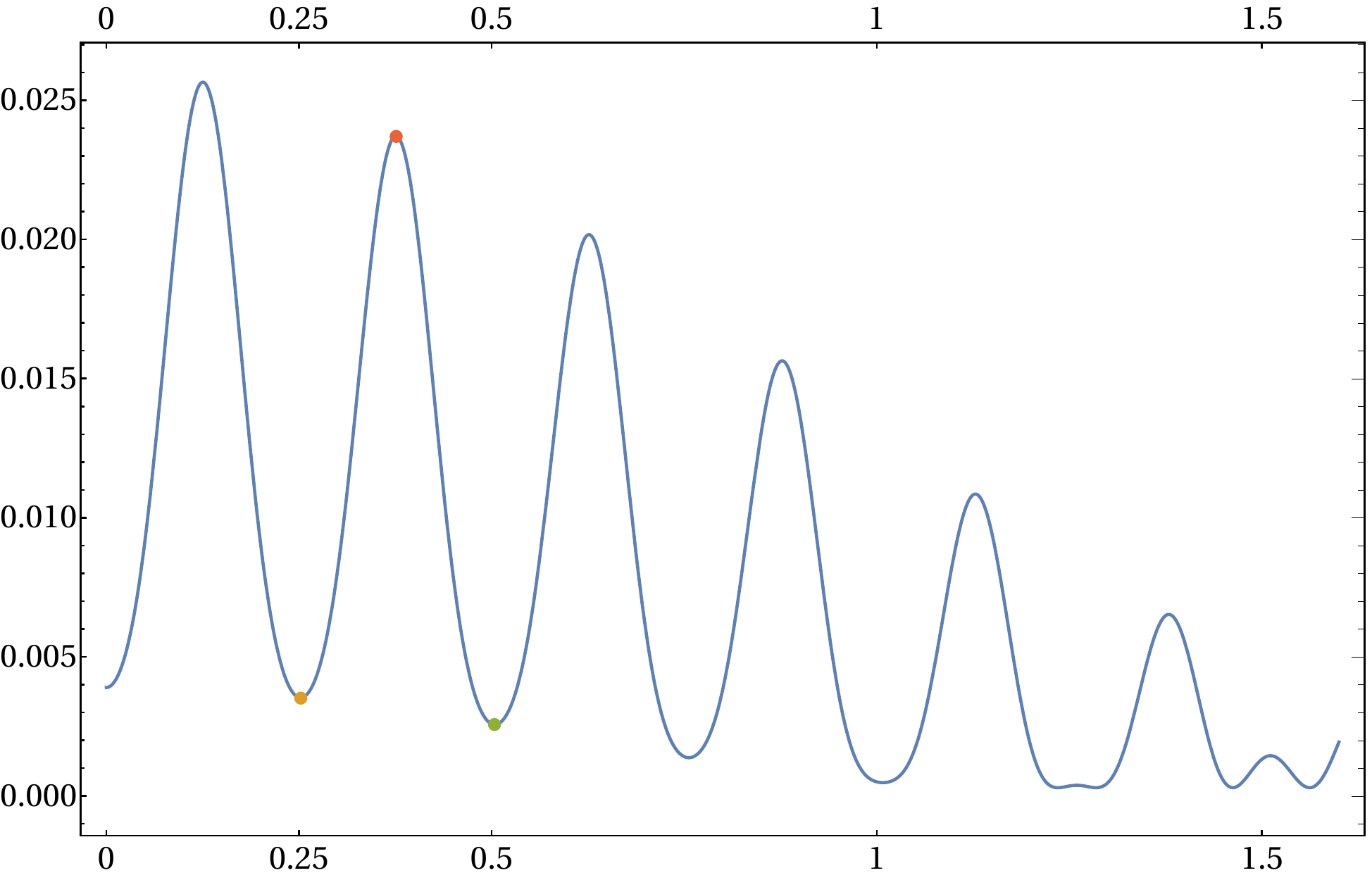}};
		\node[below= of img, node distance=0cm,yshift=1cm] {$\varphi_2$};
		\node[below= of img, node distance=0cm,xshift=-1.3cm,yshift=3cm] {$V_t$};
		\node[below= of img, node distance=0cm,xshift=-3cm,yshift=3.1cm] {$V_f$};
		\node[below= of img, node distance=0cm,xshift=-2.2cm,yshift=8.3cm] {$V_B$};
		\node[left=of img, node distance=0cm,rotate=90,anchor=center,yshift=-0.7cm] {$V_{\text{tot}}(5\pi,\varphi_2)$};
		\end{tikzpicture}
		\caption{We show the profile for the potential of Figure~\ref{fig:PotTot} at $\varphi_1=5\pi$. The orange and the green dots correspond to the values of the potential at the two local minima, respectively at $\varphi_2\sim 0.25$ and $\varphi_2\sim 0.50$. The red dot is the value of the potential at the local maximum, i.e. $\varphi_2\sim 0.38$. This plot is given for $f=1$.}
		\label{fig:ProfileVtot}
	\end{subfigure}
	\begin{subfigure}[t]{0.5\textwidth}
	    \centering
	    \begin{tikzpicture}
		\node (img) {\includegraphics[width=0.90\textwidth,keepaspectratio]{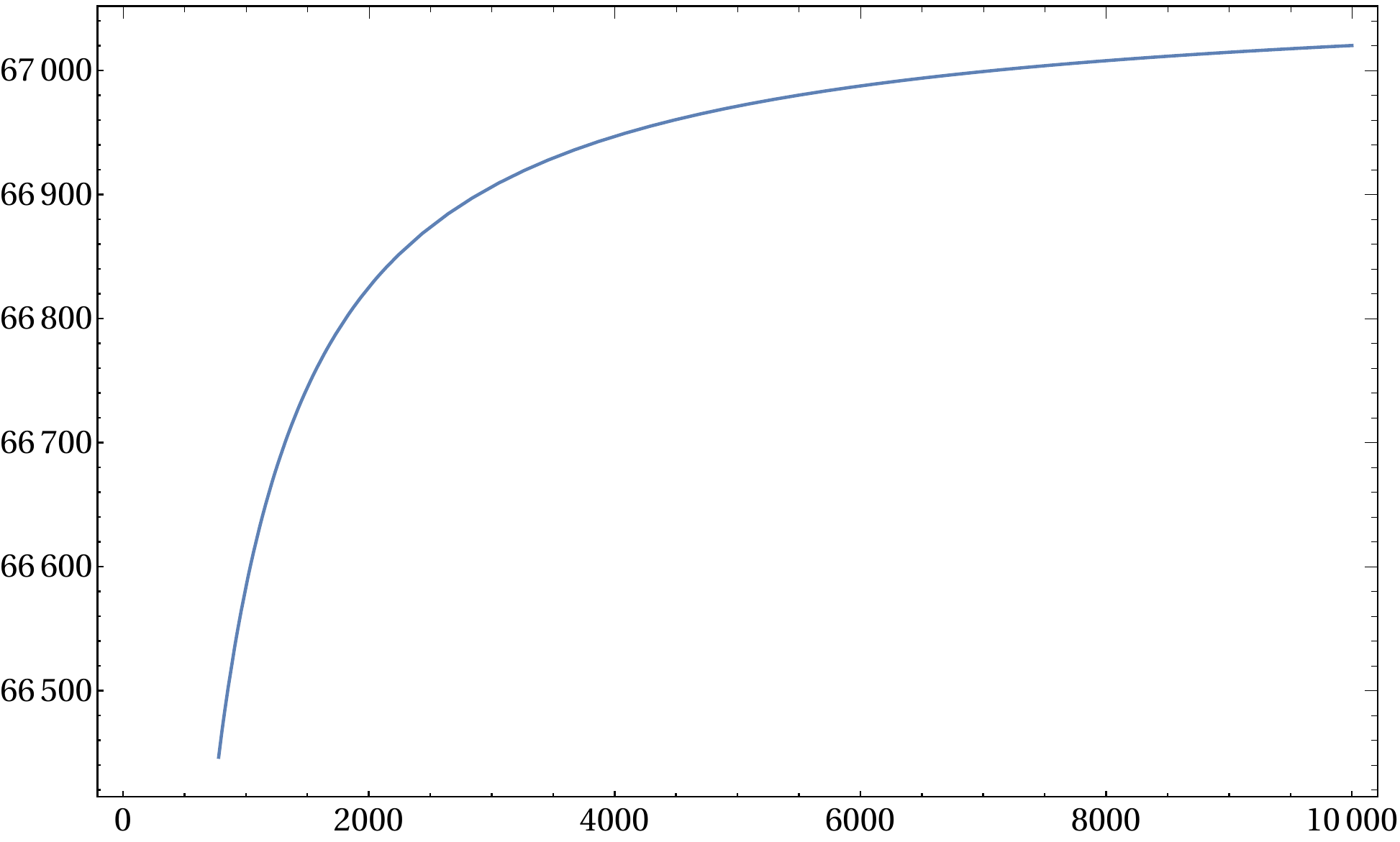}};
		\node[below= of img, node distance=0cm,yshift=1cm] {$x$};
		\node[left=of img, node distance=0cm,xshift=1.2cm] {$B$};
		\end{tikzpicture}
		\caption{$B$ defined in~\eqref{eq:BandGamma} as a function of $x$.}
		\label{fig:Bofx}
	\end{subfigure}\hfill
	\begin{subfigure}[t]{0.5\textwidth}
	    \centering
		\begin{tikzpicture}
		\node (img) {\includegraphics[width=0.90\textwidth,keepaspectratio]{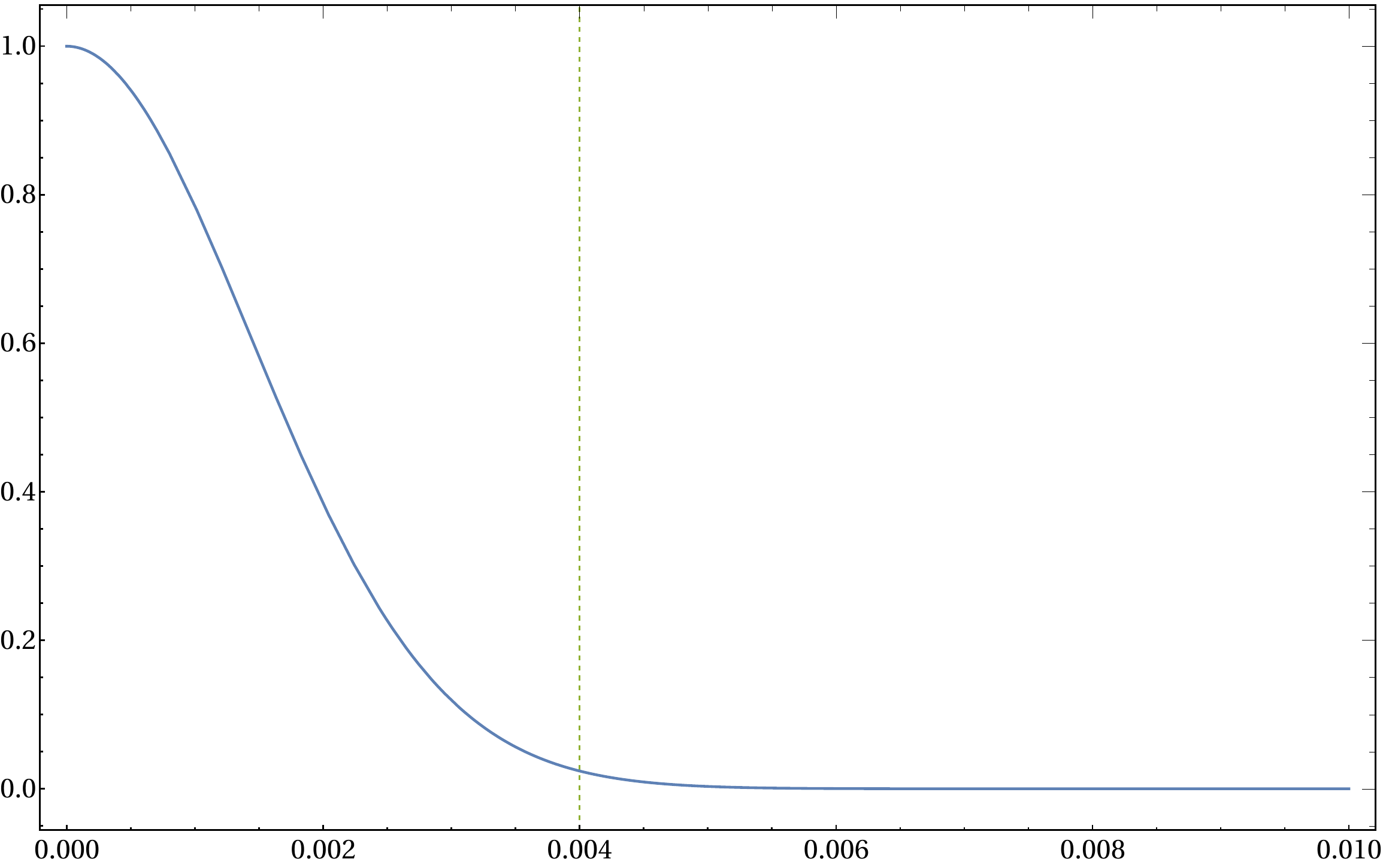}};
		\node[below= of img, node distance=0cm,yshift=1cm] {$x$};
		\node[left=of img, node distance=0cm,xshift=1.2cm] {$\Gamma$};
		\end{tikzpicture}
		\caption{$\Gamma$ defined in~\eqref{eq:BandGamma} as a function of $x$.}
		\label{fig:Gammaofx}
	\end{subfigure}
	\caption{In Figure~\ref{fig:ProfileVtot} we show the potential at the fixed value of $\varphi_1=5\pi$. We notice that since $V_B-V_f\gg V_f-V_t$, the thin-wall approximation can be used to compute the domain wall tension $T$. From Figures~\ref{fig:Bofx} and~\ref{fig:Gammaofx} we can find a critical value of $x\sim 0.004$ for which the tunneling probability is enough suppressed.}
	\label{fig:tunneling}
\end{figure}

A natural question one can ask when looking at Figure~\ref{fig:PotTot} is how likely it is for the two axions to undergo a tunneling transition between two local minima~\cite{Coleman:1980aw} of the potential $V_{tot}$. To avoid complications coming from considering a Coleman-de Luccia tunneling~\cite{Coleman:1980aw} with two fields, we restricted ourselves to compute the probability for the field $\varphi_2$ to undergo tunneling, for a fixed value of $\varphi_1$. Indeed, we set $\varphi_1$ to the value where the largest probability of tunneling is expected, i.e. on the plane where Eq.~\eqref{eq:PotTot} has a local maximum for $\varphi_1$. This happens for $\varphi_1=5\pi+10n\pi$, with $n\in \ZZ$. Looking at the sections of the potential at fixed $\varphi_1$ we can apply the well-known formulas for the decay probability for a single field~\cite{Coleman:1980aw}:\footnote{We follow the notation of~\cite{Hebecker:2020aqr}.}
\begin{equation}
    \Gamma=\exp(-B) \,\,\text{ with }\,\, B=B_0\, r(x,y)\equiv \left(\frac{27\pi^2T^4}{2(\Delta V)^3}\right)r(x,y) \fstop
    \label{eq:BandGamma}
\end{equation}
Here $B$ is the bounce action and $T$ is the tension of the domain wall. We have also defined the field theoretic bounce $B_0$ and its gravitational correction
\begin{equation}
    r(x,y)=2\frac{1+xy-\sqrt{1+2xy+x^2}}{x^2(y^2-1)\sqrt{1+2xy+x^2}}\coma
\end{equation}
with
\begin{equation}
    x=\frac{3T^2}{4M_P^2\Delta V} \coma y=\frac{V_f+V_t}{\Delta V} \,\text{ and }\, \Delta V=V_f-V_t\fstop
    \label{eq:xydeltaV}
\end{equation}
We have denoted the values of the potential in the false and true vacuum, respectively, with $V_f\equiv V_{\text{tot}}(5\pi,\varphi_2=\varphi_f)$ and $V_t\equiv V_{\text{tot}}(5\pi,\varphi_2=\varphi_t)$. 

In particular, we choose $\varphi_f\sim 0.25$ and $\varphi_t\sim 0.50$, for the plots shown in Figure~\ref{fig:tunneling}. It is clear from those that the decay rate is highly suppressed for a value of $x\gtrsim 0.004$. An important comment is now due: so far, we have not canonically normalized the kinetic term for the $\varphi_2$ axion. By doing so, the canonically normalized field space distance between the true vacuum $\varphi_t$ and the false vacuum $\varphi_f$ will depend on the axion decay constant $f$ for the $\varphi_2$ field.
We define then
\begin{equation}
    \Delta \Phi = (\varphi_t-\varphi_f)f \sim 0.25 f\fstop
\end{equation}
We further call the difference of the potential between the red and green dots in Figure~\ref{fig:ProfileVtot} as $\Delta V_B=V_B-V_f$, i.e. $\Delta V_B$ is the height of the barrier between the two minima. Since $\Delta V_B\gg \Delta V$, the thin-wall approximation is well justified in our context. In this approximation, the tension of the domain wall reads
\begin{equation}
    T=\int_{\Phi_t}^{\Phi_f} d\Phi \sqrt{2\left(V_{\text{tot}}(5\pi,\Phi/f)-V_{\text{tot}}(5\pi,\Phi_f/f)\right)} \sim \sqrt{2\Delta V_B} \,\Delta \Phi\sim 0.35 f \sqrt{\Delta V_B}\fstop
\end{equation}
From the definition of $x$ in Eq.~\eqref{eq:xydeltaV}, we can find a parametric dependence between $x$ and $f$, i.e.
\begin{equation}
    x \sim \frac{f^2}{M_P^2}\frac{\Delta V}{V_t}\fstop
\end{equation}
In order for the tunneling probability to be sufficiently suppressed, we require $B$ to be larger than an order $\mathcal{O}(100)$ number. 

In a full model with moduli stabilization consistent with an inflationary sector producing the right CMB-scale curvature perturbation, the typical scale of moduli and inflationary scalar potential will be fixed for large-field models where the slow-roll parameter is $\epsilon_V \sim 0.01$ to be $\left|V_{\rm eff.}^{\rm valley}\right|\sim 10^{-10}$. Rescaling the scalar potential in Figure~\ref{fig:ProfileVtot} to these values and reevaluating the bounce action, we get
\begin{equation}
B \sim 10^2 \left(\frac{f}{M_{\rm P}}\right)^2 \frac{1}{V_t}\fstop
\end{equation}
The longevity requirement $B\gtrsim 100$ thus translates in a lower bound on $f$, given by
\begin{equation}
    \frac{f}{M_P}\gtrsim \sqrt{V_t} \gtrsim 10^{-5}\fstop
\end{equation}
\\
In~\cref{sec:InflationGV,sec:UpliftGV} we found the conditions that the GV invariants and the VEVs of the moduli must satisfy to get, respectively, the potential for the inflationary period and for the uplift to a dS vacuum. In this section, we have introduced two parameters, i.e., \cref{eq:ep1Infl,eq:ep2def}, that must be smaller than $1$, but they must also satisfy the following relation:\footnote{Remember that we are not considering a specific CICY with $\tilde{h}^{1,1}=4$, the labels for the degrees are used only to distinguish the various GV invariants.}
\begin{equation}
    n_{1,0,0,0}e^{-\im(u_1)}\ll n_{0,1,0,0}e^{- \im(v_1)}\equiv \epsilon_1 \ll \epsilon_2 \equiv n_{0,0,1,0}e^{-\im(v_2)}\sim n_{0,0,0,1}e^{- \im(u_2)}\fstop
    \label{eq:ep1ep2cond}
\end{equation}
Following hypothesis of Section~\ref{sec:InflationGV}, we impose
\begin{equation}
    \im(v_1)\sim \im(u_1)\coma
\end{equation}
provided that 
\begin{equation}
    n_{1,0,0,0} \ll n_{0,1,0,0}\fstop
\end{equation}
One possibility is that the four saxions are all tuned to have comparable VEVs, i.e.
\begin{equation}
    \im(v_1)\sim \im(u_1)\sim \im(v_2)\sim \im(u_2)\fstop
\end{equation}
The condition~\eqref{eq:ep1ep2cond} is only satisfied for a mirror CICY with
\begin{equation}
    n_{1,0,0,0}\ll n_{0,1,0,0} \ll n_{0,0,1,0} \sim n_{0,0,0,1}\,\fstop
    \label{eq:possibility1}
\end{equation}

We used our database of all favorable CICYs with $\tilde{h}^{1,1}=4$ to see if it was possible to realize Eq.~\eqref{eq:possibility1}. The positive GV invariants\footnote{There are no negative GV invariants of degree $1$ but they could be zero.} have been ordered from the smallest to the largest and we asked a hierarchy of a factor $1:30$ or $1:10$ among three of them and the largest of three to be comparable with a fourth one with a ratio of at most $4:5$. There are no CICYs that satisfy that condition.

The other possibility is to generate the hierarchy required in Eq.~\eqref{eq:ep1ep2cond} by tuning the VEVs for $\im(v_2)$ and $\im(u_2)$ when their corresponding GV invariants are not comparable among each other but the conditions of $\epsilon_2$ still satisfy the hypothesis of Section~\ref{sec:UpliftGV}. The only other condition that Eq.~\eqref{eq:ep1ep2cond} imposes is on the VEVs between, for instance, $\im(v_2)$ and $\im(v_1)$, i.e.
\begin{equation}
    \im(v_2)-\im(v_1) \ll \ln \left(\frac{n_{0,1,0,0}}{n_{0,0,1,0}}\right)\fstop
    \label{eq:possibility2}
\end{equation}
We conclude that Eq.~\eqref{eq:possibility2} is the only possibility, and the different choices of CICYs could just change how large the VEVs must be chosen in order to satisfy~\cref{eq:possibility2}.

A generalization of this proposal to a higher number of c.s. moduli should be done completely by tuning the VEVs of the moduli that are not involved in the model. We have commented in Footnote~\ref{foot:h11decrease}, that the ratios between the smallest and the largest GV invariants reduce when $\tilde{h}^{1,1}$ increases. This means that it is always more difficult to create a hierarchy between them. We found very few CICYs that satisfy the hierarchies we are looking for to get the uplift and none of them have the correct hierarchy to make the inflationary setup. The only possibility is to look at the flux landscape and tune the VEVs of the moduli accordingly.

Given that the combined sector providing a mechanism for both inflation and uplifting works along the same lines as the individual mechanisms discussed in the previous sections, we would like to stress again that for the mirror CICYs for which our GV invariants describe the non-perturbative corrections to the c.s. moduli prepotential, a full embedding into a scenario with moduli stabilization is difficult because the required negative Euler characteristic needed for the LVS setup is absent and rigidifying all typically dozens of $4$-cycles of the mirror CICYs required to operate KKLT is difficult (for more details, please refer to the discussion in the introduction as well the last paragraph of Section~\ref{sec:UpliftGV}).

\section{Discussion and Conclusions}
\label{sec:Conclusions}

The importance of the GV invariants for phenomenological applications became evident in the previous sections. We quantified the influence of the GV invariants among the parameters involved in the construction of the inflationary model proposed in~\cite{Hebecker:2015rya} and of the uplift model in~\cite{Hebecker:2020ejb}. Interestingly, there exist CYs with hierarchies among the lowest-degree GV invariants. We explained how we can use these hierarchies to alleviate the need to tune hierarchies in the c.s. moduli.\\
In particular for the inflationary model, we found that our setup still satisfies the no-go theorem for aligned winding trajectories with two moduli proposed in~\cite{Hebecker:2018fln}.\footnote{We thank A. Hebecker for pointing out the no-go theorem to us and suggesting to check if it was still satisfied.} The issues found in~\cite{Hebecker:2018fln} for obtaining a superplanckian decay constant are still present in our construction, even if we can avoid a hierarchy among the VEVs of the moduli.\\
Additionally, using both the GV hierarchies and flux-tunable c.s moduli VEV hierarchies we present a mechanism involving a sector of four c.s. moduli which can realize both vacua with SUSY breaking and positive vacuum energy contribution and (in absence of the no-go theorem) large-field inflation. Upon combination with a proper CY realizing full moduli stabilization in an AdS vacuum, this may lead to the construction of dS vacua with an inflationary sector in type IIB string theory. While the no-go theorem still presents obstacles for this type of setup which uses on two out of four axions to arrange for inflation, we use the relative simplicity of this setup to show that the dS vacuum sector operates rather decoupled from the inflaton sector. This in turn makes it plausible that extending the inflaton sector to e.g. 3 axions to avoid the no-go theorem can still co-exist with the dS sector. We leave for future work the task of working out a full model along these lines. 

Moreover, we cannot use the CYs in the CICY database to exhibit such a full model, as their mirror symmetry partners for which we can construct the c.s. moduli sector realizing our combined mechanisms have properties which render LVS constructions impossible (the mirror-CICYs have positive Euler characteristic) and KKLT-like constructions practically difficult ($h^{1,1}$ is large).  However, if, for instance, in the future the existence of so-called Greene-Plesser mirror CY pairs were established to be widespread in the set of CICYs or, e.g. the Kreuzer-Skarke set of anticanonical hypersurfaces in toric ambient spaces, then for such pairs involving mirror partners with $h^{1,1}\geq 2$ the Greene-Plesser construction properties would likely allow realization of our combined dS and inflation sector in a full model with moduli stabilization. In anticipation of such examples, we estimated the life-time of the inflationary valleys in our combined mechanism due to Coleman-de Luccia tunneling to neighboring valleys. Quite interestingly, guaranteeing sufficient longevity of a given inflationary valley places a lower bound on the axion decay constant of the axion direction responsible for generating the valley structure of $f \gtrsim 10^{-5} M_{\rm P}$.

This application of the GV invariants of the CICYs to string phenomenology convinced us of the necessity of a database of the principal GV invariants of the CICYs up to a certain degree of the curves. We believe that such a database can be useful also for purely mathematical reasons to understand the distribution of these numbers. It would be interesting to analyze how these numbers change with respect to, for instance, $\tilde{h}^{1,1}$ of the CICYs at a fixed or also varying degree.

For interested readers and to give access to the database, we provide a \href{https://www.desy.de/~westphal/GV_CICY_webpage/GVInvariants.html}{website} to download it. We explain how to extract the data from the database in Appendix~\ref{sec:INSTANTON}, together with comments on some empirical properties of the GV invariants that we noticed.

The study of the GV invariants for the CICYs made us also look for redundancies in the CICY database. The kind of redundancies we looked for involved only a permutation of the basis elements of $H^4$ of a given CICY and we explain them in Appendix~\ref{sec:CICYreview}. We also list the tuples of CICY that have been found redundant under this kind of transformation in Table~\ref{tab:redundancies} in Appendix~\ref{sec:Tableredundancies}. It would be interesting to see if there are more redundancies and how they are distributed with respect to $\tilde{h}^{1,1}$ on the same footing of what we show in Figure~\ref{fig:RedunCICY}.

\section*{\hfil Acknowledgments \hfil}
We are grateful to A. Hebecker and J. Moritz for useful comments and discussion during the development of this project. A. M. thanks Emilio Ambite for the help with the HYDRA cluster in the IFT of Madrid, fundamental for the computations in our paper. A. M. received funding from ``la Caixa" Foundation (ID 100010434) with fellowship code LCF/BQ/IN18/11660045 and from the European Union’s Horizon 2020 research and innovation programme under the Marie Sk\l odowska-Curie grant agreement No. 713673. N. R. is supported by the Deutsche Forschungsgemeinschaft under Germany's Excellence Strategy - EXC 2121 ``Quantum Universe'' - 390833306. F.C. is supported by STFC consolidated grant ST/T000708/1.  A. W. is supported by the ERC Consolidator Grant STRINGFLATION under the HORIZON 2020 grant agreement no.  647995.

\appendix

\section{CICY Redundancies}
\label{sec:CICYreview}

In this appendix we firstly review some relevant facts about the database of complete intersection Calabi-Yau manifolds in an ambient space $\tilde{\mathcal{A}}$ given by a product of projective spaces $\PP^{n_1}\!\times ... \times\PP^{n_s}$. We later discuss the systematic search we performed, in order to check which CICYs are actually redundant, in the sense that they are topologically equivalent.

Given the ambient space $\tilde{\mathcal{A}}$, a compact K\"ahler $3$-fold can be constructed as the zero-locus of $k$ homogeneous polynomials $p_j \left(z\right)$ in $\tilde{\mathcal{A}}$, subject to the constraint:
\begin{equation}
\sum_{i=1}^{s} n_i - k = 3\fstop
\end{equation}
Each $p_j$ is characterized by its multi-degree $q_j^i$ (where $ j=1, \dots ,k $ and $ i=1,\dots ,s $), which specifies the degree in the homogeneous coordinates of each $\PP^{n_i}$. A convenient way to encode this information is by means of a configuration matrix:
\begin{equation}
\left[
\begin{tabular}{c|cccc}
$\PP^{n_1}$ &   $q_1^1$ & $\cdots$  & $q_k^1 $ \\
$\PP^{n_2}$  &   $q_1^2$ & $\cdots$  &$ q_k^2$  \\
$\vdots$ &   $\vdots$ & $\ddots$ & $\vdots$  \\ 
$\PP^{n_s} $&   $q_1^s $& $\cdots$ & $q_k^s $
\end{tabular}
\right] \fstop
\label{eq:configuration}
\end{equation}
If we require the zero-locus of the $p_j$ to be a Calabi-Yau manifold, the vanishing condition for the first Chern class imposes
\begin{equation} \label{eq:CYpolynomialcondgen}
n_i+1=\sum_{j=1}^{k} q_j^i \quad \forall \, i=1,...s \fstop
\end{equation}

A natural question that can be asked is when two Calabi-Yau manifolds are the same. In this paper, every time we say that two Calabi-Yau are the same, we mean that they are diffeomorphic as real manifolds. A famous theorem by Wall~\cite{WALL1966} implies that two simply-connected, closed Calabi-Yau 3-folds $X$ and $Y$ are isomorphic as real manifolds, if
\begin{enumerate}
    \item The Hodge numbers agree, namely $h^{1,1}(X)$=$h^{1,1}(Y)$ and $h^{2,1}(X)$=$h^{2,1}(Y)$.
    \item There exist a choice of base in $H^4(X,\mathbb{Z})$ given by $D_i$, $i=1,\ldots h^{1,1}(X)$, and a choice of base in $H^4(Y,\mathbb{Z})$ given by $\hat{D}_i$, $i=1,\ldots h^{1,1}(Y)$ such that $\int_{D_i} c_2(X)=\int_{\hat{D}_i} c_2(Y)$, where $c_2(X)$ (resp $c_2(Y)$) is the second Chern class of (the tangent bundle) of $X$ (resp $Y$).
    \item With the same choice of base of the point above for $H^4(X,\mathbb{Z})$ and $H^4(Y,\mathbb{Z})$ the triple intersection numbers agree, namely $\int_X D_i\cdot D_j\cdot D_k=\int_Y \hat{D}_i\cdot \hat{D}_j\cdot \hat{D}_k, \quad \forall i,j,k=1,\ldots h^{1,1}(X)=h^{1,1}(Y)$.
\end{enumerate}

Clearly, if two real manifolds are diffeomorphic, then this implies that also they will be homeomorphic as topological spaces, therefore topologically equivalent.

It is worth stressing that the choice of a configuration matrix for a given CICY $\tilde{X}$ is not unique, in the sense of Wall's theorem stated above. The same CY manifold $\tilde{X}$ can be realized in multiple ways by different configuration matrices. Nevertheless, different choices of the configuration matrix for the same CICY $\tilde{X}$ can make more explicit (or hide) different features of the CY itself. For example, the number of complex structure deformations visible as versal deformations of the polynomial equations, and the fibrations trivially visible from the configuration matrices, both depend on the choice of the configuration matrix for $\tilde{X}$. 

One could naively think that the construction outlined above leads to infinitely many topologically distinct CYs, as one could in principle increase both the number of $\mathbb{P}^{n_i}$ factors and their dimensions, and add more equations accordingly. However, this is false. It was shown~\cite{Green:1986ck} that all topologically distinct CYs realizable with this construction can be obtained from ambient spaces for which both the number $s$ of $\mathbb{P}^{n_i}$ factors and the size of the $n_i$ is bounded from above. Therefore, the full set of topologically distinct CICYs can be obtained from a set of finitely many configuration matrices. A database of 7890 configuration matrices was famously built in \cite{Candelas:1987kf} and it was shown that such a database is complete, in the sense that any other configuration matrix not present in the database will describe a CY topologically equivalent to the one already present in the list. We will refer to such a database as ``the old CICY database", or sometimes as ``the original CICY database".

A configuration matrix $M(\tilde{X})$ representing a CY $\tilde{X}$ for which $h^{1,1}(\tilde{X})=\tilde{h}^{1,1}(\tilde{\mathcal{A}})$ is said to be favorable. When this happens, all divisors of the CICY $\tilde{X}$ are inherited from the ones of the ambient space $\tilde{\mathcal{A}}$. It turns out that not all 7890 configuration matrices in the old CICY database are favorable, just 4896 of them are. However, favorability is not an intrinsic property of the CY $\tilde{X}$ itself, but rather depends on the choice of the configuration matrix used to describe $\tilde{X}$.

In the work of \cite{Anderson:2017aux} the old CICY database was improved: for almost all non-favorable configuration matrices in the old CICY database, a new configuration matrix representing the same CY was found, such that the new configuration matrix is now favorable. This was achieved by chains of ineffective splittings, performed on the old configuration matrix~\cite{Anderson:2017aux}. The number of favorable configurations was then pushed up to 7842. The remaining CICYs, which still does not admit a favorable configuration matrix, admit nevertheless a completely different description as a single hypersurface in a product of two del Pezzo surfaces, $\dP_m\times \dP_n$ and a theorem by Koll\'ar~\cite{Kollar:2012pv} guarantees that such description is favorable. Furthermore, out of the 7842 favorable CICYs, 22 of them are either 6-tori, or direct products of K$3$ and $2$-tori. A new database was created by keeping only the 7820 favorable and non-product CICYs. We refer to this database as ``the new CICY database", and we will use this database everywhere in our paper.

On the one hand, the new CICY database, despite being maximally favorabilized, is still not a list of unique Calabi-Yau manifolds. It is therefore important to check for redundancies, to provide for a minimal list of topologically distinct and favorable CICYs. On the other hand, the existence of redundancies in the original CICY database was realized many years ago~\cite{Candelas:1990pi,Avram:1995pu,Candelas:1989ug,Candelas:2007ac}, and many of them were identified in~\cite{Anderson:2008uw}, within the subset of the 4896 favorable CICYs of the old database. In such case, the check of the redundancies was done using Wall's theorem~\cite{WALL1966}: the authors of \cite{Anderson:2008uw} checked whether given two CICYs $\tilde{X}$ and $\tilde{Y}$ with identical Hodge numbers, they could find a change of basis in $H^4(\tilde{X},\mathbb{Z})$ and $H^4(\tilde{Y},\mathbb{Z})$ such that also the second Chern classes and triple intersection numbers agree. In particular, they focused on the change of variables given by permutations of the divisors.\footnote{An alternative way to select redundant CICYs was also proposed in~\cite{He:1990pg}. There, not only permutations of the basis elements of $H^4(\tilde{X},\mathbb{Z})$ and $H^4(\tilde{Y},\mathbb{Z})$ were considered, but also linear transformations with rational coefficients. This allowed the authors to claim the existence of some other redundancies, by finding a suitable new basis for $H^4(\tilde{X},\mathbb{Q})$ and $H^4(\tilde{Y},\mathbb{Q})$, which would now match the triple intersection number and second Chern class. However, it is not clear to us why Wall's theorem immediately applies in this case. For this reason, we decided to stick to linear changes of basis with integer coefficients, and only work with integral cohomology. Even less generally, we restrict ourselves to looking for permutations of the divisors.\label{foot:Candelas}}

We perform a similar scan within the new CICY database.
Given two CICYs with different configuration matrices, the first trivial check is to look at their Hodge numbers. If they agree, we can check if a permutation of the basis elements of $H^4(\tilde{X},\mathbb{Z})$ could exist, such that the second Chern class and the triple intersection numbers of $\tilde X$ computed in the new basis agree with those of $\tilde Y$. 
We find that there are three qualitatively distinct cases:
\begin{enumerate}
	\item \label{case1} The CICYs that already have Hodge numbers equal, (the integrals of) $c_2$ (over the base elements of $H^4$) equal, and also the intersection numbers equal. No change of basis is needed, and Wall's theorem trivially applies.
	\item \label{case2} The CICYs that have all Hodge numbers equal and (the integrals of) $c_2$ (over the base elements of $H^4$) equal. The triple intersection numbers can be made equal with a permutation of the basis elements of $H^4$ that leaves (the integrals of) $c_2$ (over the basis elements of $H^4$) unchanged. 
	\item \label{case3} Finally, the CICYs that have only Hodge numbers equal, but both (the integrals of) $c_2$ (over the basis elements of $H^4$) and the intersection numbers can be made equal with a permutation of the basis elements of $H^4$. These are the most general set. 
\end{enumerate}
We list the tuples of redundant CICYs, divided by $\tilde{h}^{1,1}$, in Appendix~\ref{sec:Tableredundancies}. In such a list, each parenthesis contains all CICYs that are redundant by a permutation of the basis elements in $H^4$.
For some of the CICYs in the cases above, we also give the explicit change of basis matrix. The list of such matrices can be accessed at \href{https://www.desy.de/~westphal/GV_CICY_webpage/GVInvariants.html}{link}.
 
 We found all redundancies up to $\tilde{h}^{1,1}=13$. We have not been able to check the most general transformations for the CICYs with $\tilde{h}^{1,1}=15$ (which are $15$) and for those with $(\tilde{h}^{1,1},\tilde{h}^{2,1})=(14,16)$ (which are $14$). However, even for $\tilde{h}^{1,1}=14,15$ we managed to find the right change of basis also for these CICYs belonging to the case~\ref{case2} above. 
  
We find around $536$ equivalence classes involving a total of $1169$ non-product favorable CICYs. This can be compared with the number of equivalence classes found in~\cite{Anderson:2008uw} and in~\cite{He:1990pg}. We find a larger number of redundancies, essentially for two reasons. Firstly, we consider the new CICY database, while in \cite{Anderson:2008uw} the authors perform this scan on the old CICY database. Since more CICYs $\tilde{X}_i$ are now favorable, it is easier to study change of basis in $H^4(\tilde{X}_i)$, since now $H^4(\tilde{X}_i)\simeq H^4(\tilde{\mathcal{A}}_i)$. Secondly, we push our scan to $\tilde{h}^{1,1}=13$ while the authors of \cite{He:1990pg} stopped at $\tilde{h}^{1,1}=6$.\footnote{Also recall the qualitative difference between our methods and those of \cite{He:1990pg}, explained in Footnote~\ref{foot:Candelas}.} Therefore we conclude that at least $6651$ CICYs are topologically distinct, and thus could lead to phenomenologically distinct models.

It is possible now to analyze the distribution of the redundant CICYs. We show in Figure~\ref{fig:RedunCICY} an histogram of the CICYs involved per $\tilde{h}^{1,1}$. The exact numbers of redundant CICYs is shown in a table next to the figure. It is interesting to compare this with the histogram of the total number of favorable non-product CICYs per $\tilde{h}^{1,1}$ in Figure~\ref{fig:allCICY}. We see that, despite the histogram in Figure~\ref{fig:allCICY} peaks at $\tilde{h}^{1,1}=7$, the redundancy histogram in Figure~\ref{fig:RedunCICY} peaks before.

\begin{figure}[htp]
    \centering
    \begin{subfigure}{\textwidth}
        \begin{minipage}{0.7\textwidth}
        \begin{scaletikzpicturetowidth}{\textwidth}
		\begin{tikzpicture}[scale=\tikzscale]
		\begin{axis}[
		ybar,
		width=\textwidth,
		bar width=16.5pt,
		enlarge y limits= 0.1,
		xlabel={$\tilde{h}^{1,1}$},
		ylabel={Redundant CICYs},
		]
		\addplot+[mark=0,draw=blue!20!black,fill=blue!30!white, opacity=0.5,every node near coord/.style={text=black},every node near coord/.append style={font=\tiny}] plot coordinates{(2, 8) (3, 50) (4, 123) (5, 190) (6, 170) (7, 155) (8, 117) (9, 114) (10, 98) (11, 72) (12, 43) (13, 17) (14, 8) (15, 5)};
		\end{axis}
		\end{tikzpicture}
		\end{scaletikzpicturetowidth}
        \end{minipage}\hfill
        \begin{minipage}[t]{0.25\textwidth}
        \begin{tabular}{c|c}
		    $\tilde{h}^{1,1}$ & \shortstack{Redundant \\CICYs } \\
		    \hline
		     1 & 0\\
		     2 & 8\\
		     3 & 50\\
		     4 & 123\\
		     5 & 190\\
		     6 & 170\\
		     7 & 155\\
		     8 & 117\\
		     9 & 114\\
		     10 & 98\\
		     11 & 72\\
		     12 & 43\\
		     13 & 17\\
		     14 & 8\\
		     15 & 5
		\end{tabular}
        \end{minipage}
	\caption{}
	\label{fig:RedunCICY}
    \end{subfigure}\\
    \begin{subfigure}{\textwidth}
    \begin{minipage}{0.7\textwidth}
        \begin{scaletikzpicturetowidth}{\textwidth}
		\begin{tikzpicture}[scale=\tikzscale]
		\begin{axis}[
		ybar,
		width=\textwidth,
		bar width=16pt,
		enlarge y limits= 0.1,
		xlabel={$\tilde{h}^{1,1}$},
		ylabel={Favorable non-product CICYs},
		]
		\addplot+[mark=0,draw=blue!20!black,fill=blue!30!white, opacity=0.5,every node near coord/.style={text=black},every node near coord/.append style={font=\tiny}] plot coordinates{(15, 15) (14, 18) (13, 68) (12, 155) (11, 368) (10, 643) (9, 1032) (8, 1325) (7, 1462) (6, 1257) (5, 856) (4, 425) (3, 155) (2, 36) (1, 5)};
		\end{axis}
		\end{tikzpicture}
		\end{scaletikzpicturetowidth}
		\end{minipage}\hfill
		\begin{minipage}[t]{0.25\textwidth}
		\begin{tabular}{c|c}
		    $\tilde{h}^{1,1}$ & \shortstack{Favorable \\CICYs } \\
		    \hline
		     1 & 5\\
		     2 & 36\\
		     3 & 155\\
		     4 & 425\\
		     5 & 856\\
		     6 & 1257\\
		     7 & 1462\\
		     8 & 1325\\
		     9 & 1032\\
		     10 & 643\\
		     11 & 368\\
		     12 & 155\\
		     13 & 68\\
		     14 & 18\\
		     15 & 15
		\end{tabular}
		\end{minipage}
	\caption{}
	\label{fig:allCICY}
    \end{subfigure}
    \caption{In Figure~\ref{fig:RedunCICY} we show the number of redundant favorable CICYs per $\tilde{h}^{1,1}$, while in Figure~\ref{fig:allCICY} we show the number of favorable non-product CICYs per $\tilde{h}^{1,1}$.}
\end{figure}

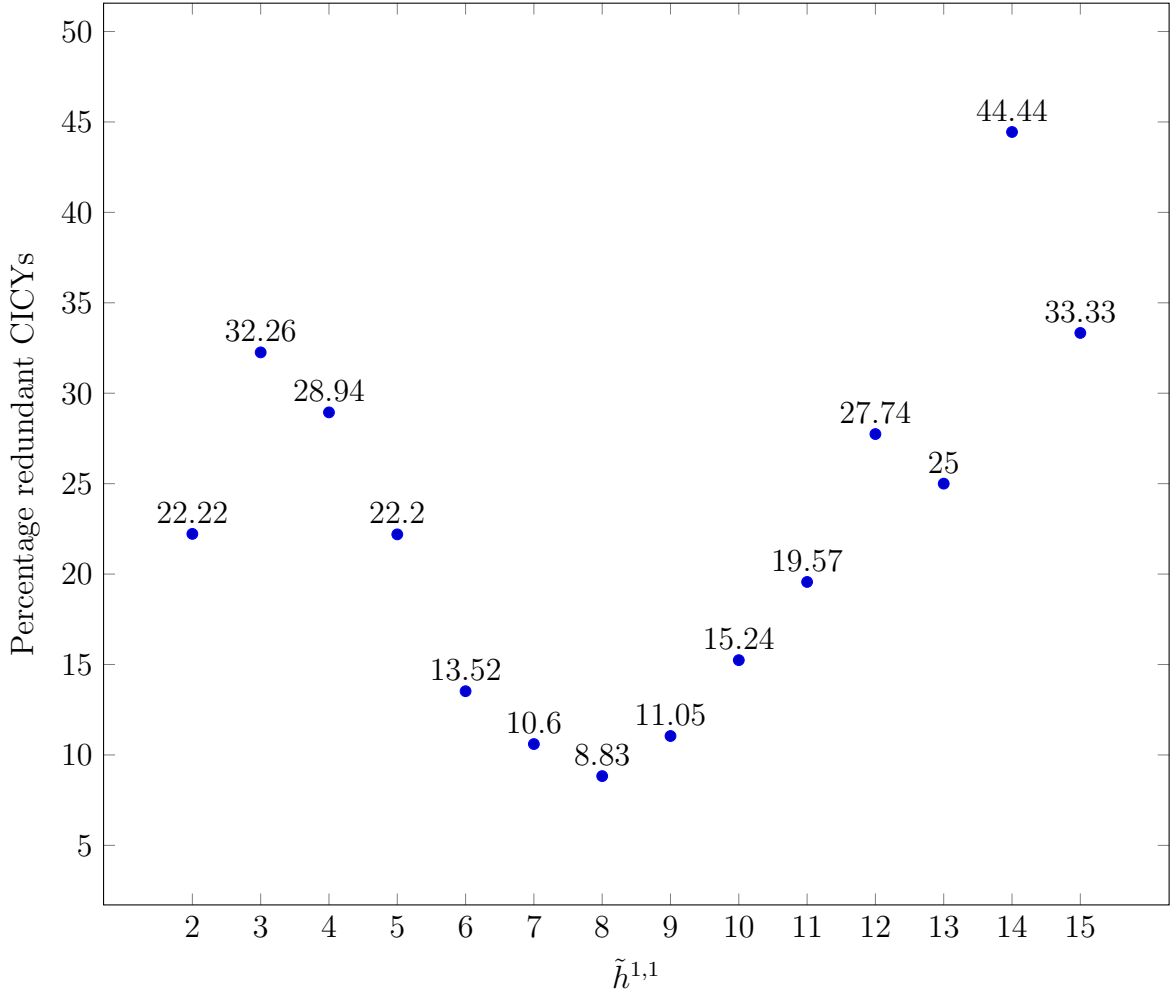
\begin{figure}[htp]
    	\begin{center}
		\begin{scaletikzpicturetowidth}{\textwidth}
		\begin{tikzpicture}[scale=\tikzscale]
		\begin{axis}[
		width=\textwidth,
		enlarge y limits= 0.2,
		nodes near coords,
        nodes near coords align={vertical},
		xtick={2,3,4,5,6,7,8,9,10,11,12,13,14,15},
        xticklabels={2,3,4,5,6,7,8,9,10,11,12,13,14,15},
        xlabel={$\tilde{h}^{1,1}$},
        ylabel={Percentage redundant CICYs},
		]
		\addplot+[only marks, mark=*, mark size=2pt,every node near coord/.style={text=black},every node near coord/.append style={font=\normalsize}] coordinates{(2., 22.2222) (3., 32.2581) (4., 28.9412) (5., 22.1963) (6.,13.5243) (7., 10.6019) (8., 8.83019) (9., 11.0465) (10.,15.2411) (11., 19.5652) (12., 27.7419) (13., 25.) (14., 44.4444) (15., 33.3333)};
		\end{axis}
		\end{tikzpicture}
		\end{scaletikzpicturetowidth}
	\end{center}
	\caption{Distribution of redundant CICYs per $\tilde{h}^{1,1}$ normalized for the number of favorable CICYs at fixed $\tilde{h}^{1,1}$.}
	\label{fig:DistRedun}
\end{figure}

Normalizing the number of redundant non-product favorable CICYs per $\tilde{h}^{1,1}$ by the number of total non-product favorable CICYs with the same $\tilde{h}^{1,1}$ we get the percentage of redundant CICYs in the plot shown in Figure~\ref{fig:DistRedun}.
It is very tempting to speculate that for some reasons the percentage number of redundant CICYs per $\tilde{h}^{1,1}$ lies on a parabola with minimum at $\tilde{h}^{1,1}=8$. This is also beautifully consistent with the following fact. Right now we are only considering redundancies in the set of favorable CICYs, however, for $\tilde{h}^{1,1}=19$ there are $15$ non-favorable CICYs which are well known to be all redundant, and all of them are the Schoen manifold~\cite{Anderson:2017aux}. Therefore, the percentage of redundant CICYs at $\tilde{h}^{1,1}=19$ is $100\%$. Adding to Figure \ref{fig:DistRedun} this extra case, we would have a point that exactly lies on the interpolating parabola found from the points in the plot.

We stress the fact that for $\tilde{h}^{1,1}=15$ we have not checked all possible combinations to find redundancies. It is possible that there are more redundant CICYs than the $5$ we have found. Looking at Figure~\ref{fig:DistRedun}, the interpolation of the shape of the distribution would suggest that there might be over $50\%$ of the CICYs with $\tilde{h}^{1,1}=15$ which are redundant. There are also $14$ CICYs with $(\tilde{h}^{1,1},\tilde{h}^{2,1})=(14,16)$ that have not been scanned completely for a generic transformation (i.e. the one belonging to the case~\ref{case3} in the previous list), but, using the same argument of the interpolation, we may expect that there are no more redundancies in that sector.

It is also possible that some more redundancies can be found by allowing for a more general linear change of base, and not just permutations. This could maybe improve the situation of points at $\tilde{h}^{1,1}=2,13,15$ in Figure \ref{fig:DistRedun}. However, it is also perfectly possible that there is no actual distribution of the redundancies and the percentage is smaller than the one naively expected by fitting the data with a parabola. 

Let us now discuss how to access the information about the change of basis matrices. For the some of the tuples collected in cases~\ref{case2} and~\ref{case3} we give the transformation matrix in a Mathematica notebook on the website \href{https://www.desy.de/~westphal/GV_CICY_webpage/GVInvariants.html}{link}. The notebook contains a table where in the first component we state on which CICY the transformation must be applied to get the other CICY. In the second component we write the transformation matrix itself. Such a matrix acts on the basis of divisors of the CICY given in the list of ~\cite{Anderson:2017aux}. We show how the matrix acts on CICYs $\{7865,7871\}$ in Appendix~\ref{sec:Tableredundancies}. For these tuples (the integrals of) $c_2$ (on the divisor basis) are trivially equal for both the CICYs, and given by
\begin{equation}
    c_2=\{24,24,56\}\fstop
\end{equation}
The intersection polynomials are naively different since
\begin{equation}
\begin{split}
    \text{R}_{7865}&=8 D_3^3+6 D_1 D_3^2+4 D_2 D_3^2+3 D_1 D_2 D_3\coma\\
    \text{R}_{7871}&=8 \tilde{D}_3^3+6 \tilde{D}_2 \tilde{D}_3^2+4 \tilde{D}_1 \tilde{D}_3^2+3 \tilde{D}_1 \tilde{D}_2 \tilde{D}_3\fstop
\end{split}
\end{equation}
In the Mathematica notebook, we give the matrix 
\begin{equation}
    M=\left(
\begin{array}{ccc}
 0 & 1 & 0 \\
 1 & 0 & 0 \\
 0 & 0 & 1 \\
\end{array}
\right)\coma 
\end{equation}
that transforms the basis $D_i$ permuting $D_1$ with $D_2$ and we notice that such matrix does not change the values of $c_2$.\\
A similar example can be done for the tuple $\{7574,7593\}$ that has the most general transformation we considered. For those CICYs we have
\begin{equation}
\begin{split}
    c_{2|7574}&=\{24,24,52,44\}\coma\\
    c_{2|7593}&=\{24,24,44,52\}
\end{split}
\end{equation}
and the intersection polynomials read
\begin{equation}
\begin{split}
    \text{R}_{7574}=&4 D_3^3+4 D_1 D_3^2+2 D_2 D_3^2+10 D_4 D_3^2+8 D_4^2 D_3+2 D_1 D_2 D_3+6 D_1 D_4 D_3+\\
    &+5 D_2 D_4 D_3+2 D_4^3+2 D_1 D_4^2+4 D_2 D_4^2+3 D_1 D_2 D_4\coma\\
    \text{R}_{7593}=&2 \tilde{D}_3^3+4 \tilde{D}_1 \tilde{D}_3^2+2 \tilde{D}_2 \tilde{D}_3^2+8 \tilde{D}_4 \tilde{D}_3^2+10 \tilde{D}_4^2 \tilde{D}_3+3 \tilde{D}_1 \tilde{D}_2 \tilde{D}_3+5 \tilde{D}_1 \tilde{D}_4 \tilde{D}_3+\\
    &+6 \tilde{D}_2 \tilde{D}_4 \tilde{D}_3+4 \tilde{D}_4^3+2 \tilde{D}_1 \tilde{D}_4^2+4 \tilde{D}_2 \tilde{D}_4^2+2 \tilde{D}_1 \tilde{D}_2 \tilde{D}_4\fstop
\end{split}
\end{equation}
In the Mathematica notebook, we give the matrix 
\begin{equation}
    M=\left(
\begin{array}{cccc}
 0 & 1 & 0 & 0 \\
 1 & 0 & 0 & 0 \\
 0 & 0 & 0 & 1 \\
 0 & 0 & 1 & 0 \\
\end{array}
\right)\coma
\end{equation}
that transforms $c_{2|7593}$ into $c_{2|7574}$ but also matches the two intersection polynomials. 

\section{A database of Gopakumar-Vafa invariants for CICYs}
\label{sec:INSTANTON}

In this appendix we recall the usual technique to compute the genus $0$ GV invariants of Calabi-Yau threefolds, as explained in \cite{Hosono:1993qy,Hosono:1994ax}. By using this technique, we created a database of GV invariants for the set of favorable complete intersection Calabi-Yau's, searching for compactification spaces showing the required hierarchy of invariants to make viable the models of~\cref{sec:InflationGV,sec:UpliftGV,sec:InflationUpliftGV}.

Suppose we want to compute the GV invariants of a given CICY $\tilde{X}$. Let $t_i$, $i=1,\ldots \tilde{h}^{1,1}$ be the number of K\"ahler moduli of such manifold. By mirror symmetry, there will exist a mirror manifold $X$ with c.s. moduli $z_i$, $i=1,\ldots,h^{2,1}=\tilde{h}^{1,1}$.\footnote{Following the convention introduced in the main text (Footnote~\ref{foot:h11h21conv}), we denote $h^{1,1}(\tilde{X})$ as $\tilde{h}^{1,1}$.} The main idea of the algorithm will be to explicitly compute the period vector in the mirror side $X$, and then from this extract the quantum corrected triple intersection numbers of the CICY $\tilde{X}$.

A configuration matrix for $\tilde{X}$ as in Eq.~\eqref{eq:configuration} is given by
\begin{equation}
	\left[
	\begin{tabular}{c|cccc}
		$\PP^{n_1}$ &   $q_1^1$ & $\cdots$  & $q_k^1 $ \\
		$\PP^{n_2}$  &   $q_1^2$ & $\cdots$  &$ q_k^2$  \\
		$\vdots$ &   $\vdots$ & $\ddots$ & $\vdots$  \\ 
		$\PP^{n_{\tilde{h}^{1,1}}} $&   $q_1^{\tilde{h}^{1,1}} $& $\cdots$ & $q_k^{\tilde{h}^{1,1}} $
	\end{tabular}
	\right] \fstop
\end{equation}

From the generators of the Mori cone of the mirror manifold $X$, it is possible to define vectors $l^{(i)}$, given by
\begin{equation}
	l^{(i)}=\left(-q_1^{(i)},\ldots -q_{k}^{(i)}\,;\, \ldots, 0,1,\ldots, 1,0,\ldots\right)\equiv\left(\left\{l_{0j}^{(i)}\right\}\,;\,\left\{l_r^{(i)}\right\}\right)\coma
\end{equation}
where $i=1,\ldots, h^{2,1}$ and $j=1,\ldots, k$ and the number of $1$'s in $\left\{l_r^{(i)}\right\}$ are equal to $n_i+1$ at a position corresponding to the $\PP^{n_i}$ that has been considered. 

The period vector $\Pi(z)$ for $X$ is a vector with $2h^{2,1}+2$ components. The first component, also called the \emph{fundamental period}, is given by
\begin{equation}
	w_0(z)=\sum_{n_1\geq 0}\ldots \sum_{n_{h^{2,1}}\geq 0}c(n) \prod_{i=1}^{h^{2,1}}z_i^{n_i}\coma
\end{equation}
where\footnote{$n!=\Gamma(n+1)$ is the Euler's Gamma function.}
\begin{equation}
	c(n)=\dfrac{\displaystyle\prod_j\Gamma\left(1-\sum_{s=1}^{h^{2,1}}l_{0j}^{(s)}n_s\right)}{\displaystyle\prod_{i}\Gamma\left(1+\sum_{s=1}^{h^{2,1}}l_i^{(s)}n_s\right)}\fstop
\end{equation}
Notice in particular that it is possible to write down the fundamental period of $X$, just from the information encoded in the configuration matrix of $\tilde{X}$.

One then extends such a solution of the Picard-Fuchs for arbitrary values of $h^{2,1}$ parameters $\rho_i$, defining
\begin{equation}
	w_0(z,\rho)=\sum_{n_1\geq 0}\ldots \sum_{n_{h^{2,1}}\geq 0}c(n+\rho) \prod_{i=1}^{h_{2,1}}z_i^{n_i+\rho_i}\fstop
	\label{eq:genfunperiod}
\end{equation}
In terms of \eqref{eq:genfunperiod}, the full period vector $\Pi(z)$ can be defined as~\cite{Hosono:1993qy,Hosono:1994ax}
\begin{equation}
	\renewcommand*{\arraystretch}{2}
	\Pi(z)=\left(\begin{array}{*3{>{\displaystyle}c}p{5cm}} w_0(z) \\ \left.\pderr{\rho_i}w_0(z,\rho)\right|_{\rho=0}\\ \left.\frac{1}{2}\kappa^0_{ijk}\pderr{\rho_j}\pderr{\rho_k}w_0(z,\rho)\right|_{\rho=0}\\
		\left.-\frac{1}{6}\kappa^0_{ijk}\pderr{\rho_i}\pderr{\rho_j}\pderr{\rho_k}w_0(z,\rho)\right|_{\rho=0}
	\end{array}\right)\coma
\end{equation}
where $\kappa_{ijk}^0$ are the classical triple intersection numbers of $\tilde{X}$. 

At this point one has obtained the GV invariants for $X$, but in order to extract them one needs to rewrite such period vector in terms of the K\"ahler moduli of $\tilde{X}$, which are defined by the mirror map
\begin{equation}
	t^i(z)=\frac{w_i(z)}{w_0(z)}\coma
	\label{eq:tinwz}
\end{equation}
where
\begin{equation}
	w_i(z)=\sum_{n_1\geq 0}\ldots \sum_{n_{h^{2,1}}\geq 0}\left.\frac{1}{2\pi i}\pderr{\rho_i}c(n+\rho)\right|_{\rho=0}\prod_{i=1}^{h^{2,1}}z_i^{n_i}+w_0(z)\frac{\ln z_i}{2\pi i}\fstop
\end{equation}
At the technical level, the most complicated point of the algorithm is the inversion of Eq.~\eqref{eq:tinwz} to get the c.s. moduli $z$ as a function of $t$. This is the part which limits the most every attempted implementation of the code.

The quantum-corrected triple intersection numbers $\kappa_{ijk}$ can be expressed as
\begin{equation}
	\kappa_{ijk}(t)=\pderr{t_i}\pderr{t_j}\dfrac{\displaystyle \left.\frac{1}{2}\kappa^0_{kab}\pderr{\rho_a}\pderr{\rho_b}w_0(z,\rho)\right|_{\rho=0}}{w_0(z)}(t)\coma
	\label{eq:kappacomp}
\end{equation}
where it is clear that the fraction is computed first as function of the c.s. moduli $z_i$, then, one substitutes the inverse of Eq.~\eqref{eq:tinwz}, and takes the last two derivatives with respect to the K\"ahler moduli $t_i$.

Let us introduce
\begin{equation}
	q_i=\exp\left(2\pi i t_i\right)\coma
\end{equation}
and the general expression for $\kappa_{ijk}$ as
\begin{equation}
	\kappa_{ijk}=\kappa_{ijk}^0+\sum_{d_1\geq 0}\ldots \sum_{d_{\tilde{h}^{1,1}}\geq 0}n_{d_1,\ldots,d_{\tilde{h}^{1,1}}}d_i d_j d_k\dfrac{\displaystyle\prod_{l=1}^{\tilde{h}^{1,1}}q_l^{d_l} }{\displaystyle 1-\prod_{l=1}^{\tilde{h}^{1,1}}q_l^{d_l}}\fstop
	\label{eq:kappagen}
\end{equation}
Matching the coefficients of the series expansion in $q_i$ for both~\cref{eq:kappacomp,eq:kappagen}, it is possible to extract the GV invariants $n_{d_1,\ldots d_{\tilde{h}^{1,1}}}$ for a given CICY.

The algorithm, schematically reviewed above, was coded in the Mathematica program \texttt{INSTANTON}~\cite{Klemm:2001aaa}. By using such a program, we collected all genus $0$ GV invariants for all the favorable CICYs listed by~\cite{Anderson:2017aux} up to $\tilde{h}^{1,1}=9$. For any CICY in this subset, we computed all GV invariants such that the sum of their degrees is smaller or equal than $10$. 

It is possible to find the list of the invariants on the website \href{https://www.desy.de/~westphal/GV_CICY_webpage/GVInvariants.html}{link}. They are divided in zip files by $\tilde{h}^{1,1}$, each one containing a \texttt{.dat} file named with the number of the CICY they are referred to, following~\cite{Anderson:2017aux}. The extraction of the GV invariants can be done with a simple pattern search. Here we provide a pseudo-code in Mathematica for that. Suppose you have put the files \texttt{.dat} on a folder with a Mathematica notebook. Then it is possible to extract all the numbers of the CICYs in the folder from the name of the files using 

\vspace{0.5cm}

 \begin{lstlisting}[language=Mathematica]
numberCICY = 
Thread[FileBaseName[FileNames["*.dat", NotebookDirectory[]]]]
\end{lstlisting}
while we can import the i$^{\text{th}}$ CICY in \texttt{numberCICY} with
 \begin{lstlisting}[language=Mathematica]
GVCICY = Import[StringJoin[numberCICY[[i]] <> ".dat"], 
"Table", FieldSeparators -> "\n"]
\end{lstlisting}
Finally, the degree of the j$^{\text{th}}$ curve and the corresponding value of the GV invariants can be found using
\begin{lstlisting}[language=Mathematica]
degree = Flatten[ToExpression[StringReplace[
Flatten[StringCases[GVCICY[[j]], 
RegularExpression["\[(.*?)\]"]]], {"[" -> "{", "]" -> "}"}]]]

value = ToExpression[StringDrop[Flatten[
StringCases[GVCICY[[j]], RegularExpression["\=(.*$)"]]], 1]]
\end{lstlisting}

\vspace{0.5cm}

We now comment on some empirical properties of the GV invariants in the database, and some patterns which we recognized.

For any given favorable CICY $\tilde{X}$ the Mori cone will be $\tilde{h}^{1,1}$ dimensional. For every integer point in the Mori cone, there corresponds a curve class $[\beta]$, and one can compute the genus $0$ GV invariants for this curve class. One can then move further away in the following sense. Pick any line passing through $[\beta]$, with rational angular coefficient. Such a line will hit the boundary of the Mori cone on one side, but will continue indefinitely towards infinity on the other side. In particular, it will intersect infinitely many integer points inside the Mori cone, each corresponding to a curve class. One can then compute the GV invariants for curve classes lying on such line. There are three qualitatively different ways in which the GV invariants behave when moving towards infinity in the Mori cone, in a specific direction. For some choices of the direction, the GV invariants will grow indefinitely and exponentially. We will call such directions \emph{exponentially infinite rays}.

Much more interesting is a second type of behavior, in which for some specific directions the GV invariants will eventually become zero. We will call these directions \emph{vanishing rays}. An important role is played by those vanishing rays which are normal to a boundary of the Mori cone. As already pointed out in~\cite{Candelas:1989js,Candelas:1989ug,Hosono:1994ax} for the CICYs and in~\cite{Demirtas:2020ffz} in the context of the Kreuzer-Skarke database, the existence of such vanishing rays signals the presence of a conifold transition, or a flop.\footnote{The existence of flop phases in CICYs was recently discovered in \cite{Brodie:2020fiq}.} In particular the GV invariants of a CY $\tilde{Y}$ connected to $\tilde{X}$ by a conifold transition can be recovered by summing all the GV invariants of $\tilde{X}$ in each of those vanishing ray. We illustrate this in the context of the CICY 7858, which is connected by a conifold transition to the quintic.

\begin{figure}[t]
    \centering
		\begin{scaletikzpicturetowidth}{\textwidth}
		\begin{tikzpicture}[scale=\tikzscale]
		\node (img) {\includegraphics[width=0.85\textwidth,keepaspectratio]{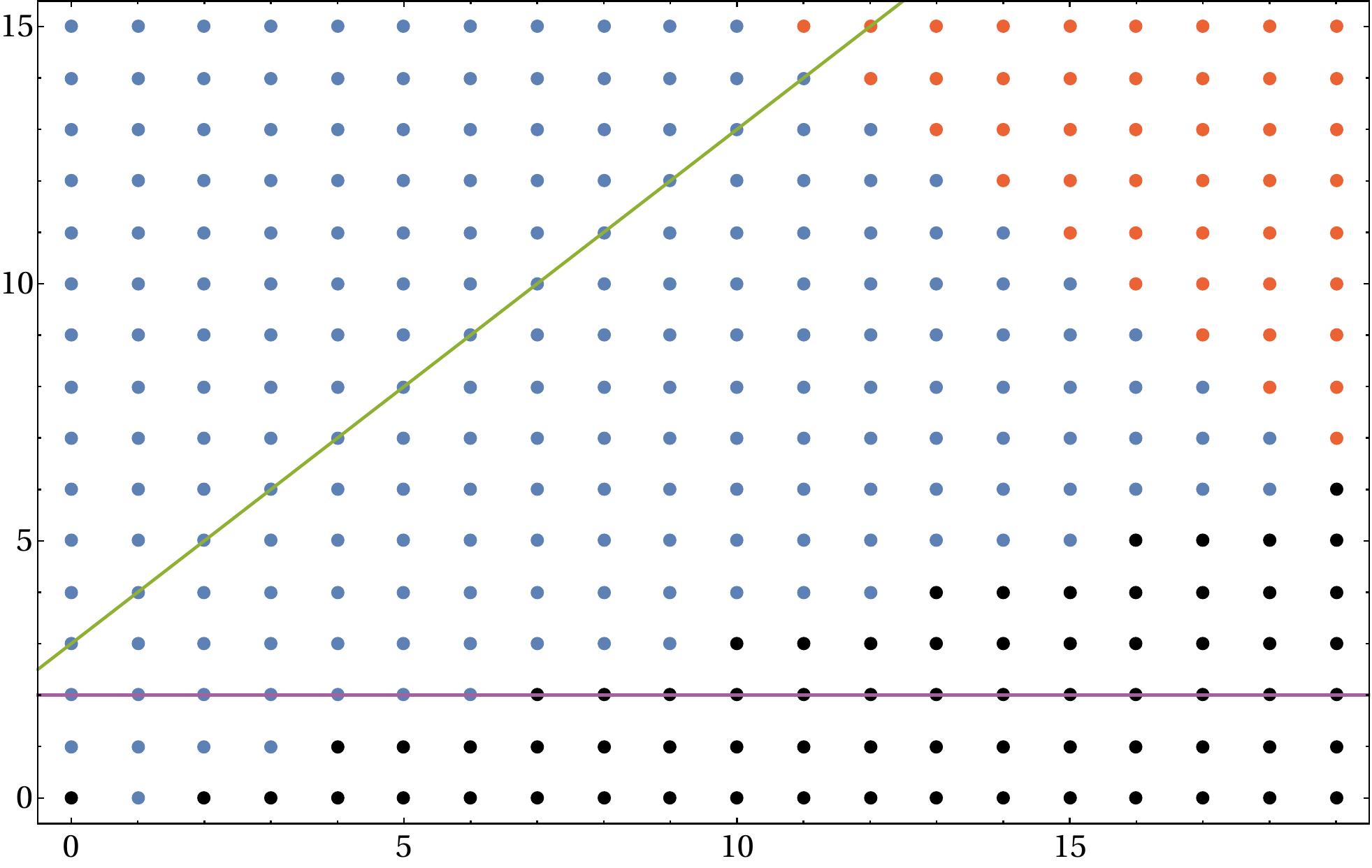}};
		\node[below= of img, node distance=0cm,yshift=1cm] {$\beta_1$};
		\node[left=of img, node distance=0cm,rotate=90,anchor=center,yshift=-0.7cm] {$\beta_2$};
		\end{tikzpicture}
		\end{scaletikzpicturetowidth}
    \caption{Occupation sites for the CICY $7858$.}
    \label{fig:GVplot}
\end{figure}

In Figure \ref{fig:GVplot} we plot the Mori cone of the CICY 7858. We put a blue dot for every curve class $[\beta]$ for which we computed that $n_{[\beta]}\neq 0$. We put a red dot for all curve classes for which we have not computed the GV invariant, but we strongly believe it is going to be non-zero. We finally put a black dot for all curve classes such that $n_{[\beta]}= 0$. We can clearly see that, for example, the ray given by $(0,3)+\mbox{Span}(1,1)$ (corresponding to the green line in Figure \ref{fig:GVplot}) is an infinite ray. On the other hand, the ray given by $(0,2)+\mbox{Span}(1,0)$ (corresponding to the purple line in Figure \ref{fig:GVplot}) is a vanishing ray. We see that in general, in this example, all rays of the form $(0,n)+\mbox{Span}(1,0)$, for all $n\in \mathbb{N}$ are vanishing rays.

The Mori cone of the quintic is then identified with the vertical axis in the figure, and the GV invariants of the quintic of degree $i$, can be found by summing over all GV invariants corresponding to the same vanishing ray normal to the boundary of the Mori cone. Namely,
\begin{equation}
n_{i}=\sum_{j=1}^{\infty} n_{j,i}\fstop
\label{eq:conifoldrelation}
\end{equation}
We can see explicitly that this is true since, for example, for the quintic $n_{1}=609250$, while the non-vanishing GV invariants on the purple vanishing ray of Figure \ref{fig:GVplot} for the CICY 7858 are
\begin{equation}
\begin{aligned}
&n_{0,2}=2670\coma n_{1,2}=73728\coma n_{2,2}=255960\coma \\ &n_{3,2}=231336\coma n_{4,2}=45216\coma n_{5,2}=360\coma n_{6,2}=-20,
\end{aligned}
\end{equation}
and we can verify that Eq.~\eqref{eq:conifoldrelation} is satisfied. The same holds for any other vanishing ray perpendicular to the boundary of the Mori cone of the quintic in Figure \ref{fig:GVplot}. Although we discussed just one specific example here, we observe that this phenomenon is generic in the CICY database and can be regarded as a confirmation of the well-known fact that all CICYs are connected by conifold transitions~\cite{Candelas:1989ug}. For every couple of CICYs connected by a single conifold transition, the GV invariants of the two manifolds are related in the manner discussed above. This behavior is expected, as, to access a conifold transition from the resolved side, one shrinks some $\mathbb{P}^1$ curves, and therefore projects the Mori cone onto one of its boundaries.

We now move to a third type of interesting direction in the Mori cone, which we call \emph{infinite periodic ray}. Along these directions, the GV invariants continue to be always non-vanishing, but they do not grow exponentially. Instead, they will repeat periodically. We observe this phenomenon, for example, in 13 $\tilde{h}^{1,1}=2$ CICYs ($7643$, $7668$, $7725$, $7758$, $7807$, $7808$, $7821$,	$7833$, $7844$, $7853$, $7868$, $7883$ and $7884$), in particular for the GV invariants $n_{0,m}$. We do not have an argument for why such periodicity arises. However, we note empirically that this is related to the presence of $\mathbb{P}^2$ factors in the ambient space geometry. A peculiar example of this is the bi-cubic CICY ($7884$), where the invariants repeat along both the $[1,0]$ and the $[0,1]$ direction of the Mori cone and are $\mathbb{Z}_2$ symmetric. One can find that infinite periodic rays also exist for $\tilde{h}^{1,1}=3$, anytime a $\mathbb{P}^3$ is present in the configuration matrix. We conjecture that this phenomenon is generic. However, as we go to a larger $\tilde{h}^{1,1}$, it is more difficult to study such behavior.

The last thing that we notice from our database is the fact that the numerical values of degree $1$ GV invariants tend to decrease with $\tilde{h}^{1,1}$. For example, the quintic has $\tilde{h}^{1,1}=1$ and its degree $1$ GV invariant is $n_1=2875$, the largest one in the whole database. On the other hand, the degree $1$ GV invariants of the CICY number $7858$ of Figure \ref{fig:GVplot} are
\begin{equation}
n_{0,1}=366\coma n_{1,0}=36\fstop
\end{equation}
We wish to address these empirical properties in a future work.

\section{List of redundancies in the CICY list}
\label{sec:Tableredundancies}
\renewcommand{\arraystretch}{1.5}
\begin{tabularx}{\textwidth}{|*{7}{c|}X|*{5}{l}p{35mm}}
	\hline
	$\tilde{h}^{1,1}$  & Tuples of redundancies \\
	\hline
	$2$ & $\{7816,7822\},\{7819,7823\},\{7867,7869\},\{7886,7888\}$ \\
	\hline
	$3$ & $\{7450,7481\},\{7464,7485\},\{7465,7466\},\{7558,7584\},\{7560,7579\},\{7570,7587\},$\\
	&  $\{7576,7577\},\{7578,7588\},\{7626,7647\},\{7627,7645\},\{7638,7648\},$\\
	&  $\{7714,7735,7745\},\{7720,7730\},\{7721,7734\},\{7753,7769\},\{7755,7763\},$\\
	&  $\{7779,7789\},\{7780,7788,7792,7795\},\{7782,7783\},\{7841,7848,7851\},$\\
	&  $\{7843,7847\},\{7865,7871\},\{7877,7881\}$\\
	\hline
	$4$ & $\{5784,5823\},\{5805,5806\},\{6206,6224\},\{6524,6558\},\{6527,6557\},\{6545,6546\},$\\
	&  $\{6776,6839\},\{6784,6828\},\{6814,6815\},\{6816,6835\},\{7039,7074\},\{7042,7113\},$\\
	&  $\{7049,7080\},\{7056,7077\},\{7059,7075\},\{7089,7090\},\{7204,7218,7241,7270\},$\\
	&  $\{7205,7245\},\{7209,7292\},\{7217,7250,7285\},\{7221,7248,7276\},$\\
	&  $\{7222,7223,7274,7275\},\{7277,7278\},\{7334,7405\},\{7348,7391\},\{7349,7373\},$\\
	&  $\{7352,7400\},\{7354,7395\},\{7355,7389\},\{7357,7394\},\{7359,7371\},\{7423,7424\},$\\
	&  $\{7435,7462,7491,7522\},\{7446,7493\},\{7449,7476,7490,7495\},\{7458,7497\},$\\
	&  $\{7461,7498\},\{7468,7507\},\{7478,7492,7505\},\{7548,7605\},\{7557,7582\},$\\
	&  $\{7569,7598\},\{7572,7573\},\{7574,7593\},\{7618,7629,7654\},\{7628,7649\},$\\
	&  $\{7639,7650\},\{7656,7657\},\{7681,7690\},\{7719,7736,7742\},\{7722,7733\},$\\
	&  $\{7751,7772\},\{7786,7793\},\{7818,7825\},\{7820,7827\}$\\
	\hline
	$5$ & $\{5249,5313\},\{5256,5301,5452\},\{5270,5338\},\{5271,5340\},$\\
	&  $\{5284,5285,5322,5323\},\{5352,5355\},\{5353,5354,5356\},\{5758,5896\},$\\
	&  $\{5775,5914\},\{5780,5835\},\{5793,5849\},\{5814,5860\},\{5821,5830\},\{6040,6041\},$\\
	&  $\{6091,6313\},\{6110,6329\},\{6121,6122\},\{6146,6278\},\{6168,6330\},\{6169,6332\},$\\
	&  $\{6176,6225\},\{6191,6323\},\{6205,6248\},\{6208,6233\},\{6209,6239\},\{6210,6235\},$\\
	&  $\{6211,6236\},\{6255,6256\},\{6520,6631\},\{6542,6577\},\{6543,6575\},\{6544,6610\},$\\
	&  $\{6554,6584\},\{6582,6583\},\{6659,6660\},\{6711,6941\},$\\
	&  $\{6715,6788,6836,6927,6947\},\{6722,6911\},\{6726,6931,6946\},$\\
	&  $\{6732,6777,6802,6834,6890,6896\},\{6734,6735,6805,6806,6899,6900\},$\\
	&  $\{6736,6737,6901,6902\},\{6741,6908\},\{6743,6808,6895\},\{6744,6914\},$\\
	&  $\{6747,6891\},\{6758,6789,6856\},\{6764,6909\},\{6766,6916\},\{6768,6926\},$\\
	&  $\{6775,6841\},\{6787,6855\},\{6794,6860\},\{6821,6843\},\{6822,6842\},\{6823,6869\},$\\
	&  $\{6825,6866\},\{7010,7092\},\{7012,7098\},\{7026,7093\},\{7028,7102\},\{7029,7107\},$\\
	&  $\{7058,7085\},\{7167,7306\},\{7174,7279\},\{7194,7225,7257,7269\},\{7195,7272\},$\\
	&  $\{7198,7289\},\{7207,7251\},\{7231,7258\},\{7287,7288\},\{7294,7295\},\{7338,7393\},$\\
	&  $\{7345,7401\},\{7442,7508\},\{7443,7511\},\{7444,7526\},\{7447,7487\},\{7474,7494\},$\\
	&  $\{7502,7503\},\{7513,7514\},\{7547,7600\},\{7625,7660\},\{7705,7741\}$\\
	\hline
	$6$ & $\{3208,3513\},\{3902,4255\},\{3905,4283\},\{4027,4178\},\{4039,4194\},\{4041,4195\},$\\
	&  $\{4055,4218\},\{4575,4924\},\{4605,4606,4798,4799\},\{4622,4787\},\{4623,4788\},$\\
	&  $\{4630,4631,4632,4633\},\{4669,4826\},\{4671,4827\},\{4681,4838\},\{4740,4795\},$\\
	&  $\{5111,5535\},\{5125,5502\},\{5165,5525\},\{5166,5526\},\{5167,5534\},\{5168,5532\},$\\
	&  $\{5177,5319\},\{5187,5336\},\{5196,5369\},\{5238,5454,5518\},\{5241,5457\},$\\
	&  $\{5281,5342\},\{5297,5377\},\{5528,5552\},\{5597,5598\},\{5667,5668\},\{5690,5937\},$\\
	&  $\{5694,5960\},\{5702,5888\},\{5703,5889\},\{5708,5948\},\{5721,5863\},\{5722,5864\},$\\
	&  $\{5723,5862\},\{5729,5859\},\{5733,5734\},\{5751,5886\},\{5753,5887\},\{5755,5885\},$\\
	&  $\{5766,5901\},\{6087,6333\},\{6089,6334\},\{6099,6302,6367\},$\\
	&  $\{6100,6101,6300,6301,6363,6364\},\{6103,6365\},\{6104,6339\},\{6135,6253\},$\\
	&  $\{6136,6252\},\{6139,6266\},\{6151,6286\},\{6152,6289\},\{6154,6290\},\{6156,6254\},$\\
	&  $\{6187,6281\},\{6202,6231\},\{6471,6646\},\{6472,6636\},\{6473,6647\},\{6474,6637\},$\\
	&  $\{6488,6590\},\{6489,6589\},\{6497,6573\},\{6511,6613\},\{6623,6625\},\{6723,6934\},$\\
	&  $\{6759,6868\},\{6761,6904\},\{6774,6840\},\{7000,7001\},\{7011,7121\},\{7014,7117\},$\\
	&  $\{7030,7031\},\{7228,7249\},\{7460,7499\}$\\
	\hline
	$7$ & $\{2317,2335,2665\},\{2354,2682\},\{2359,2688\},\{2375,2701\},\{2820,2871\},$\\
	&  $\{2821,2857\},\{3193,3351\},\{3194,3352\},\{3202,3374\},\{3224,3524\},\{3225,3526\},$\\
	&  $\{3226,3528\},\{3353,3498\},\{3828,4315\},\{3829,4314\},\{3836,4038\},\{3987,4348\},$\\
	&  $\{3994,4147\},\{3995,4146\},\{4029,4250\},\{4030,4268\},\{4044,4202\},\{4071,4108\},$\\
	&  $\{4073,4117\},\{4078,4185\},\{4096,4097\},\{4139,4140\},\{4225,4232\},\{4243,4342\},$\\
	&  $\{4244,4336\},\{4245,4329\},\{4264,4341\},\{4274,4346\},\{4349,4350,4351\},$\\
	&  $\{4530,4620\},\{4537,4573\},\{4549,4572\},\{4550,4571\},\{4551,4579\},\{4649,4819\},$\\
	&  $\{4748,4761\},\{4963,4964\},\{5102,5216\},\{5104,5105\},\{5109,5213\},$\\
	&  $\{5141,5259,5406\},\{5144,5519\},\{5145,5497\},\{5150,5485,5548\},\{5197,5531\},$\\
	&  $\{5214,5520\},\{5215,5512\},\{5223,5493\},\{5226,5445\},\{5228,5447\},\{5252,5310\},$\\
	&  $\{5272,5396\},\{5277,5311,5423\},\{5278,5422\},\{5280,5341\},\{5424,5449\},$\\
	&  $\{5484,5547\},\{5749,5943\},\{5750,5944\},\{5781,5837\},\{5880,5906,5917\},$\\
	&  $\{6141,6343\},\{6159,6353\},\{6161,6162\},\{6165,6324\},\{6192,6241\},\{6599,6622\},$\\
	&  $\{6807,6865\},\{7265,7282,7299\}$\\
	\hline
	$8$ & $\{1620,1908\},\{1627,1920\},\{1689,2064\},\{1702,1910\},\{2284,2803\},\{2322,2683\},$\\
	&  $\{2323,2684\},\{2324,2685\},\{2336,2666\},\{2380,2381\},\{2480,2631\},\{2512,2787\},$\\
	&  $\{2513,2786\},\{2529,2662\},\{2577,2617,2659\},\{2785,2892\},\{3114,3622\},$\\
	&  $\{3120,3621\},\{3141,3590\},\{3164,3307\},\{3165,3306\},\{3213,3245\},\{3299,3300\},$\\
	&  $\{3396,3434\},\{3397,3436\},\{3398,3419\},\{3435,3484,3499\},\{3453,3454\},$\\
	&  $\{3456,3457\},\{3476,3503\},\{3890,4337\},\{3891,4324\},\{3898,3911\},\{3910,3956\},$\\
	&  $\{3927,4087,4198\},\{3928,4199\},\{3939,4227\},\{4019,4359\},\{4043,4201\},$\\
	&  $\{4051,4210\},\{4103,4164\},\{4135,4208\},\{4172,4209,4219\},\{4174,4216\},$\\
	&  $\{4263,4313\},\{4524,4525\},\{4674,4828\},\{4753,4767\},\{4884,4944\},$\\
	&  $\{5210,5240,5409,5455\},\{5227,5446\},\{5350,5387,5414\},\{5844,5913\},$\\
	&  $\{6082,6123\},\{6857,6922\}$\\
	\hline
	$9$ & $\{1003,1312\},\{1102,1463\},\{1121,1247\},\{1122,1160\},\{1138,1309\},\{1158,1231\},$\\
	&  $\{1159,1325\},\{1285,1300\},\{1666,2039\},\{1667,2040\},\{1695,1797\},\{1703,1790\},$\\
	&  $\{1705,1706\},\{1724,1806\},\{1789,1879\},\{1791,1880\},\{1832,1864,1873\},$\\
	&  $\{1846,1849\},\{2334,2338,2366,2663,2670,2699\},\{2337,2667\},\{2339,2534\},$\\
	&  $\{2350,2477\},\{2357,2403,2596,2640\},\{2361,2383,2645,2660,2834,2839\},$\\
	&  $\{2363,2689\},\{2364,2503\},\{2376,2394\},\{2389,2465\},\{2426,2594\},$\\
	&  $\{2427,2514,2595,2647\},\{2445,2446\},\{2447,2448\},\{2481,2632\},\{2505,2506\},$\\
	&  $\{2748,2865\},\{2811,2823,2825\},\{2818,2826\},\{3207,3379\},\{3260,3330\},$\\
	&  $\{3295,3363\},\{3329,3375\},\{3918,4197\},\{3945,4024\},\{3966,4168\},\{4133,4220\},$\\
	&  $\{4262,4292\},\{4586,4678\},\{5115,5239\},\{5182,5183\},\{5184,5237\}$\\
	\hline
	$10$ & $\{577,886\},\{660,749\},\{664,688\},\{667,670\},\{683,684\},\{726,729\},\{735,827\},$\\
	&  $\{736,759\},\{737,758,814\},\{748,839\},\{754,836\},\{756,835\},\{974,975\},$\\
	&  $\{1116,1250\},\{1139,1147,1243,1310,1313,1402\},\{1163,1164\},\{1166,1169\},$\\
	&  $\{1181,1220\},\{1184,1219\},\{1191,1193\},\{1192,1194\},\{1244,1403\},\{1307,1407\},$\\
	&  $\{1315,1316\},\{1328,1428\},\{1330,1430,1435\},\{1333,1381,1424\},$\\
	&  $\{1427,1441\},\{1452,1453,1454\},\{1911,2031\},\{1912,1921,1933\},\{1925,2058\},$\\
	&  $\{1934,2057\},\{1971,1972\},\{1987,2049\},\{2032,2033\},\{2047,2048\},\{2449,2450\},$\\
	&  $\{2591,2646\},\{2629,2661\},\{2664,2807\},\{2743,2794\},\{2765,2795,2804\},$\\
	&  $\{3531,3591\}$\\
	\hline
	$11$ & $\{264,322\},\{269,318\},\{270,282\},\{271,281,315\},\{276,327,414,481\},\{278,324\},$\\
	&  $\{279,323\},\{341,372\},\{369,370\},\{383,403\},\{386,388\},\{412,426,428\},$\\
	&  $\{419,445\},\{420,442\},\{425,427\},\{440,467\},\{446,447\},\{450,453\},\{572,653\},$\\
	&  $\{742,763,765\},\{764,766\},\{768,811,832\},\{772,805,818\},\{773,806\},\{778,780\},$\\
	&  $\{779,781\},\{1004,1008,1021\},\{1010,1095\},\{1051,1087\},\{1286,1295\},$\\
	&  $\{1342,1343\},\{1827,1883\}$\\
	\hline
	$12$ & $\{143,152\},\{144,153,179\},\{155,168\},\{157,159\},\{158,160\},\{165,166\},$\\
	&  $\{170,172\},\{171,173\},\{175,176\},\{190,204\},\{194,195\},\{209,213,223\},$\\
	&  $\{220,224\},\{222,226\},\{273,284,285\},\{311,317\},\{355,356\},\{661,701\},$\\
	&  $\{1176,1249\},\{2641,2837\}$\\
	\hline
	$13$ & $\{75,77\},\{82,87\},\{83,84\},\{102,109\},\{119,121\},\{197,198\},\{415,465,484\},$\\
	&  $\{466,480\}$\\
	\hline
	$14$ & $\{56,57\},\{59,60\},\{103,104\},\{107,111\}$\\
	\hline
	$15$ & $\{3,12\},\{7,11,13\}$\\
	\hline
\caption{List of redundancies in the CICY list.}
\label{tab:redundancies}
\end{tabularx}

\newpage
\bibliographystyle{JHEP}
\bibliography{RefCicy}

\end{document}